\documentclass[11pt]{article}

\usepackage{arxiv}

\usepackage[utf8]{inputenc} 
\usepackage[T1]{fontenc}    
\usepackage{hyperref}       
\usepackage{url}            
\usepackage{amsfonts}       
\usepackage{amsmath,amssymb,mathtools}
\usepackage{microtype}      
\usepackage{cleveref}       
\usepackage{graphicx,color}

\newcommand {\beq} {\begin{equation}}

\newcommand {\eeq} {\end{equation}}
\newcommand {\beqa} {\begin{eqnarray*}}
	\newcommand {\eeqa} {\end{eqnarray*}}

\newcommand {\bls} {\boldsymbol}

\newcommand{\beqs}{\begin{eqnarray}}
\newcommand{\eeqs}{\end{eqnarray}}
\newcommand{\half}{\frac{1}{2}}

\newcommand{\bfsigma}{\boldsymbol {\sigma}}
\newcommand{\bfepsilon}{\boldsymbol {\epsilon}}

\newcommand{\bfomega}{\boldsymbol {\omega}}

\newcommand{\bftau}{\boldsymbol {\tau}}

\newcommand{\trace}{\mathop{\rm tr}\nolimits}

\newcommand{\bfe}{{\boldsymbol e}}

\newcommand{\bfu}{{\boldsymbol u}}

\newcommand{\bfA}{{\boldsymbol A}}
\newcommand{\bfB}{{\boldsymbol B}}
\newcommand{\bfC}{{\boldsymbol C}}
\newcommand{\bfD}{{\boldsymbol D}}
\newcommand{\bfE}{{\boldsymbol E}}
\newcommand{\bfF}{{\boldsymbol F}}
\newcommand{\bfG}{{\boldsymbol G}}
\newcommand{\bfH}{{\boldsymbol H}}
\newcommand{\bfI}{{\boldsymbol I}}

\newcommand{\bfM}{{\boldsymbol M}}

\newcommand{\bfP}{{\boldsymbol P}}
\newcommand{\bfQ}{{\boldsymbol Q}}

\newcommand{\edit}[1]{{\color{black}{#1}}}
\renewcommand{\vec}[1]{\boldsymbol{#1}}

\title{Representing the stress and strain energy of elastic solids with initial stress and transverse texture anisotropy}

\author{
Soumya Mukherjee \\
Department of Mechanical Engineering \\
National Institute of Technology Jamshedpur \\ Jamshedpur, Jharkhand 831014, India.\\
\texttt{2019rsme015@nitjsr.ac.in}
\And
Michel Destrade \\
School of Mathematical and Statistical Sciences\\
University of Galway\\ Galway, Ireland \\
Department of Engineering Mechanics\\ Zhejiang University\\
Hangzhou 310027, PR China
\And
Artur L. Gower \\
Department of Mechanical Engineering \\
The University of Sheffield\\  Sheffield, United Kingdom\\
\texttt{arturgower@gmail.com} 
}

\renewcommand{\shorttitle}{Elastic solids with initial stress and texture anisotropy}

\begin{document}

\maketitle


\begin{abstract}
Real-world solids, such as rocks, soft tissues, and engineering materials, are often under some form of stress. Most real materials are also, to some degree, anisotropic due to their microstructure, a characteristic often called the `texture anisotropy'. This anisotropy can stem from preferential grain alignment in polycrystalline materials, aligned micro-cracks, or structural reinforcement, such as collagen bundles in biological tissues, steel rods in prestressed concrete and reinforcing fibres in composites.
Here we establish a framework for initially stressed solids with transverse texture anisotropy.
We consider that the strain energy per unit mass of the reference is an explicit function of the elastic deformation gradient, the initial stress tensor, and the texture anisotropy.
We determine the corresponding constitutive relations and develop examples of nonlinear strain energies which depend explicitly on the initial stress and direction of texture anisotropy.
As an application, we then employ these models to analyse the stress distribution of an inflated initially stressed cylinder with texture anisotropy, and the tension of a welded metal plate.
We also deduce the elastic moduli needed to describe linear elasticity from stress reference with transverse texture anisotropy. As an example we show how to measure the stress with small-amplitude shear waves.
\end{abstract}


\section{Introduction}

\begin{figure}[h!]
\centering
\includegraphics[width=\textwidth]{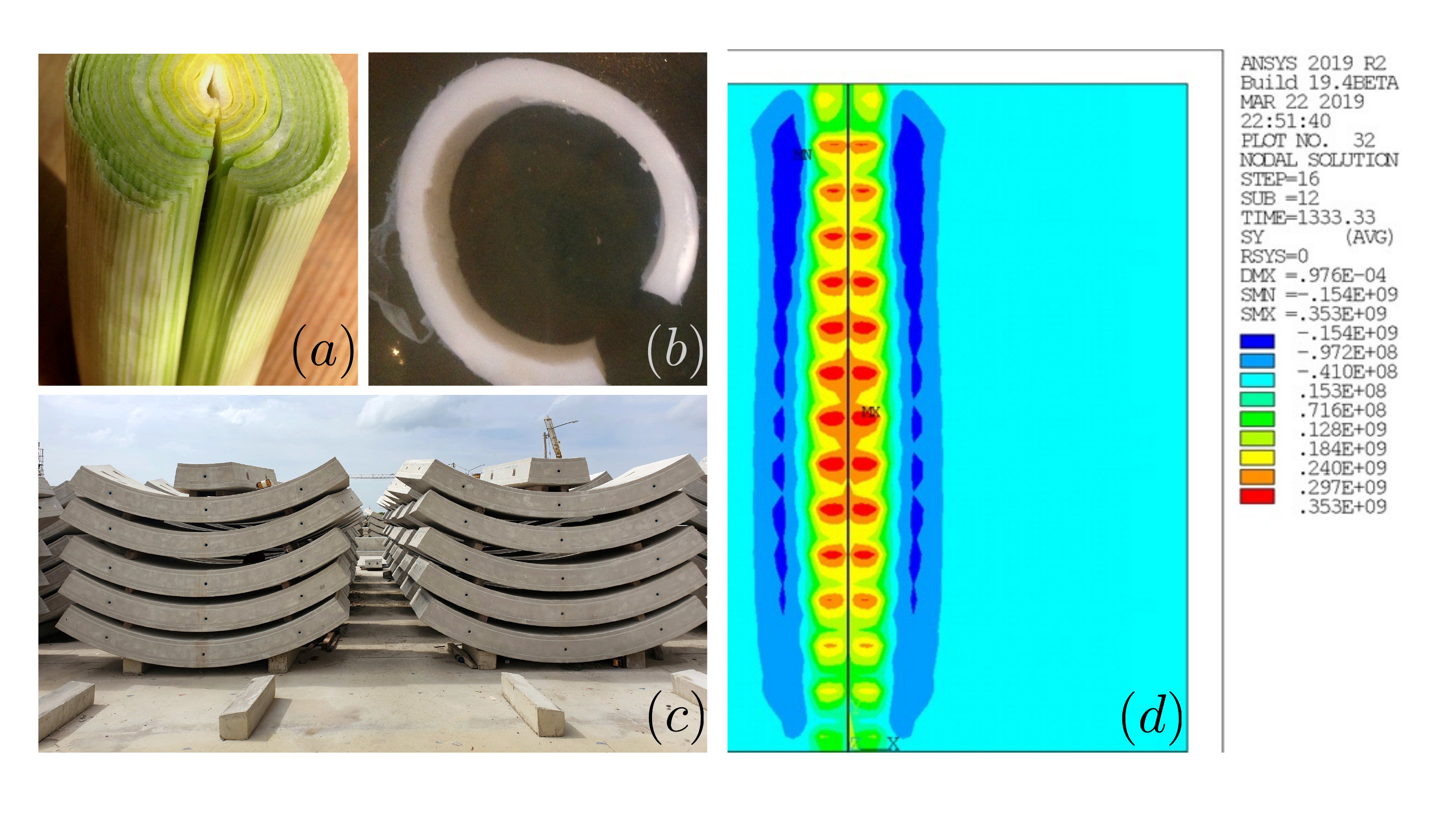}
\caption{Residual stresses and texture anisotropy are present in many natural and man-made structures.
A piece of leek (a)  or a ring of squid (b)  both spontaneously open up when cut radially, revealing that they were subject to circumferential residual stresses (the leek clearly has a microstructure aligned with its axis which creates texture anisotropy in that direction).
Concrete slabs (c) are pre-stressed by metallic rods to ensure they are under compression everywhere (shutterstock.com), with the rods creating texture anisotropy.
Finite Element simulations (d), \edit{made with Ansys® Academic Research Mechanical, } show that high temperature metal welding creates very high compressive thermal stresses that can cause compressive plastic deformation localised at the weld zone. This plastic deformation introduces high tensile and compressive residual stresses at the weld-zone and away from it.
This build-up is characteristic of many manufacturing processes such as metal cutting, rolling, machining, welding, etc. }
\label{fig:1}
\end{figure}

There are many materials with texture anisotropy that are under stress in their natural state, such as metals, rocks and other polycrystalline materials, or biological soft and hard tissues~\cite{vandiver2009differential,taber2001stress}, see examples in Figure \ref{fig:1}.
It is crucial to account for both  initial stress and texture when designing testing methods for these materials, especially non-destructive inspection methods based on the propagation of ultrasonic elastic waves~\cite{thompson1986angular, kube2015acoustoelasticity, man1996separation, thompson1986angular, li2020ultrasonic}.
To understand the elastic response of these materials and link it directly to the initial stress, we must derive constitutive equations from a strain energy function which explicitly depends on both initial stress and texture anisotropy~\cite{gower2015initial}.

In this paper, we deduce strain energy functions $W$ and necessary restrictions on the Cauchy stress $\bfsigma$, when they are explicit functions of the initial stress and texture anisotropy. We define initial stress as the stress in some reference configuration, which can have any origin. When this stress is present without any traction applied at the boundary, it is  known as residual stress.


Hoger and collaborators \cite{hoger1985residual,hoger1993constitutive,hoger1994positive,hoger1996elasticity,johnson1993dependence,johnson1995use} developed the first models for initially stressed hyperelastic materials.
Johnson and Hoger \cite{johnson1993dependence} presented the idea of a virtual stress-free configuration, obtained by cutting an initially stressed body into infinitely many small pieces and releasing all the initial stress stored. They showed numerically that the virtual stress-free configuration can be used to appropriately model the material in an initially stressed configuration, as well.

Strain energy functions can be written explicitly  in terms of the initial stress by considering the combined invariants of the initial stress with the elastic deformation gradient, see the work of Ogden and collaborators \cite{shams2011initial, ogden2011propagation, merodio2013influence}. Gower and collaborators~\cite{gower2015initial,gower2017new} then introduced a necessary restriction for the strain energy and stress, when these depend on the reference only through the initial stress (and mass density), and elastic deformation gradient; they called the restriction ISRI (Initial Stress Reference Independence).
They showed that without this restriction, absurd results could arise, even for a uniaxial deformation, see the example given in the beginning of \cite{gower2017new}.
These models have since led to practical new methods to measure stress~\cite{li2020ultrasonic}.
In this paper, we also use ISRI, applying it now to initially stressed materials that also have texture anisotropy.

Gower \textit{et al.} \cite{gower2017new} developed two models for initially stressed compressible isotropic materials satisfying ISRI of the neo-Hookean type; Agosti \textit{et al.} \cite{agosti2018constitutive} proposed incompressible Mooney-Rivlin models; Mukherjee \cite{mukherjee2022constitutive, mukherjee2022influence} determined a model for incompressible stressed Gent materials \cite{mukherjee2022constitutive}, and obtained a failure model in the presence of residual stress \cite{mukherjee2022influence}.
These models helped to examine how growth and initial stress are coupled~\cite{du2018modified, du2019influence, du2019prescribing, liu2020growth}, to study wave propagation~\cite{ciarletta2016residual} and wrinkles/creases~\cite{ciarletta2016morphology} in residually stressed tubes, and to investigate the static and dynamic characteristics of composite spheres \cite{mukherjee2021static}.
In this paper, we develop two strain energy functions  for initially stressed materials that are compressible and have texture anisotropy, and demonstrate their practical use by
solving a boundary value problem involving the inflation of a thick-walled cylinder.

Among other relevant works, we mention those by Merodio et al.~\cite{merodio2016extension} and Shariff et al.\cite{shariff2017spectral} for initially stressed isotropic materials, and by Ogden and Singh \cite{ogden2011propagation} and Shariff \cite{shariff2021anisotropic} for initially stressed structurally anisotropic materials, some of which do not satisfy ISRI \cite{gower2015initial,gower2017new}.

Finally we note that linearised theories for initially stressed isotropic materials provide a powerful tool for many small-strain problems.
This type of theory would prove most useful for designing various experiments, like elastic wave propagation, to measure initial stress.
Indeed, cutting an initially stressed solid into an infinite number of pieces to arrive at a stress-free configuration is not practical in the real world, so that ultrasonic non-destructive evaluation might be the only avenue available to testing. Grine \cite{grine2021initially} employed linear constitutive relations for stressed isotropic materials to numerically determine the mechanical fields in a cracked body. So far, there seems to be no appropriate linearised theory for initially stressed transversely isotropic materials available.

This paper is organised as follows. In Section \ref{isrisec}, we formulate initial stress reference independence for initially stressed transversely isotropic materials. We employ the reference independence and other properties of initial stress to determine the invariants here. These invariants are then used to derive constitutive relations. In Section \ref{sec:example}, we develop some reference-independent strain energy functions for initially stressed compressible transversely isotropic materials by using
initial stress symmetry (ISS)~\cite{gower2015initial}.
With one of the models, we solve \edit{two boundary value problems: inflation of initially stressed tubes and stretching of a welded joint}.
Finally in Section \ref{Lin_} we  determine the linearised constitutive relation for small strain and small initial strain, and employ ISRI to put restrictions on the material parameters. \edit{We show how the resulting equations can be used to design a method to measure the initial stress with shear waves.}


\section{Initial stress reference independence for textural transverse isotropy}
\label{isrisec}


In this section, we develop the principle of reference independence for initially stressed transversely isotropic materials and establish other properties of initial stress to determine the required constitutive relations, including the  invariants required for hyperelastic modelling.

The constitutive relation for elastic solids can be determined from the strain energy density $W$ per unit volume of reference.
For a material with initial stress tensor $\bftau$, and with a preferred direction of textural symmetry along a vector $\bfM$, the strain energy density has the functional dependence $W=W\left(\bfF,\bftau,\bfM\right)$, where $\bfF$ is the elastic deformation gradient from the reference configuration $\mathcal{R}$ to the current configuration $\mathcal{C}$.

\begin{figure}[!htbp]
    \centering
    \includegraphics[scale=1.05]{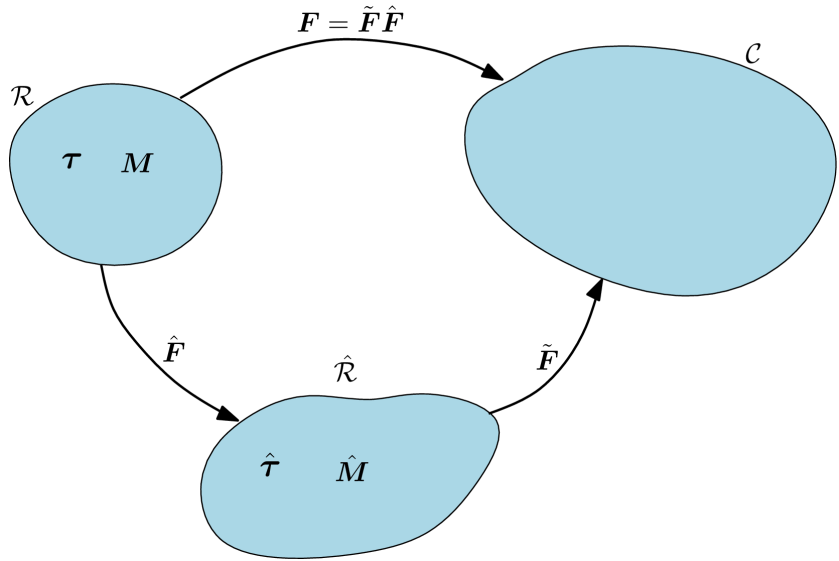}
    \caption{The three configurations for initially stressed solids: (a) the reference configuration $\mathcal{R}$ with initial stress $\bftau$ and texture direction $\bfM$ (b) the reference configuration $\hat{\mathcal{R}}$ with initial stress $\hat{\bftau}$ and texture direction $\hat{\bfM}$ (c) the current configuration $\mathcal{C}$.}
    \label{fig:isri}
\end{figure}

Figure \ref{fig:isri} depicts three configurations ($\mathcal{R}$, $\hat{\mathcal{R}}$ and $\mathcal{C}$) for initially stressed solids with textural symmetry.
We consider that the deformation (with gradient $\bfF$) taking place between $\mathcal{R}$ and $\mathcal{C}$ may equivalently by decomposed into a deformation (with gradient $\hat{\bfF}$) from $\mathcal{R}$ to $\hat{\mathcal{R}}$, an intermediate configuration, followed by  another deformation (with gradient $\tilde{\bfF}$) from $\hat{\mathcal{R}}$ to $\mathcal{C}$.
As a result, $\bfF = \tilde{\bfF}\hat{\bfF}$ and $J = \tilde J \hat J$, where $J = \det\bfF$, $\tilde J = \det \tilde\bfF$, $\hat J = \det \hat \bfF$.
The preferred directions of texture anisotropy are given by $\bfM$ and $\hat{\bfM}$ in the configurations $\mathcal{R}$ and $\hat{\mathcal{R}}$, respectively, and related by
\begin{equation}
\hat{\bfM}=\hat{\bfF}\bfM,
\label{eqn:texture-reln}
\end{equation}
noticing that $\hat{\bfM}$ is not necessarily a unit vector.
We call $\bftau$ and $\hat{\bftau}$ the initial stress fields in configurations $\mathcal{R}$ and $\hat{\mathcal{R}}$, respectively.

We now enforce the principle of Initial Stress Reference Independence (ISRI) \cite{gower2017new}, stating that the energy density of the configuration $\mathcal C$ should be the same whether it is arrived at by the direct deformation from ${\mathcal{R}}$ or the composed deformation via $\hat{\mathcal{R}}$.
This translates as
\begin{equation}
W(\bfF, \bftau, \bfM) = \hat J W\left(\tilde {\bfF}, \boldsymbol \sigma(\hat \bfF, \bftau, \bf M), \hat \bfM\right),
\end{equation}
where $\boldsymbol \sigma(\hat \bfF, \bftau, \bf M)$ is the Cauchy stress in $\hat{\mathcal R}$.
We call it $\hat{\bftau}$, and rewrite this identity as
\begin{equation}
    J^{-1}W\left(\bfF,\bftau,\bfM\right)=\tilde{J}^{-1}W\left(\tilde{\bfF},\hat{\bftau},\hat{\bfM}\right),
    \label{eqn:1st}
\end{equation}
holding for all $\bftau$, $\bfM$, $\tilde{\bfF}$, and $\hat{\bfF}$.
Differentiating both sides of \eqref{eqn:1st} with respect to  $\tilde{\bfF}$, we obtain, after some algebra,
\begin{equation}
    J^{-1}\bfF\frac{\partial W}{\partial \bfF}\left(\bfF,\bftau,\bfM\right)=\tilde{J}^{-1}\tilde{\bfF}\frac{\partial W}{\partial \tilde{\bfF}}\left(\tilde{\bfF},\hat{\bftau},\hat{\bfM}\right),
\end{equation}
which provides ISRI in terms of Cauchy stress in the current configuration $\mathcal{C}$ as follows:
\begin{equation} \label{eqn:stress-ISRI}
    \bfsigma\left(\bfF,\bftau,\bfM\right)=\bfsigma\left(\tilde{\bfF},\hat{\bftau},\hat{\bfM}\right).
\end{equation}

Now we investigate the dependence of stress and energy density over their arguments.

Because  any rigid body rotation (represented by the proper orthogonal tensor $\bfQ$, say) should not result in a change of energy density, we write that
$
W\left(\bfF,\bftau,\bfM\right)=W\left(\bfQ\bfF,\bftau,{\bfM}\right)$, which ensures that $W$ depends on $\bfF$ through the right Cauchy-Green deformation tensor $\bfC = \bfF^T\bfF$.
Moreover, we assume that the energy density is the same when reversing the direction of $\bfM$ to $-\bfM$. Consequently, the energy density should depend directly on the \emph{texture tensor} $\bfG=\bfM\otimes\bfM$.
As a result, $W = W\left(\bfF,\bftau,\bfM\right)$ can also be seen as a function of $\bfC$, $\bftau$, and $\bfG$,
which enables the following representation of stress,
\begin{equation}
    \bfsigma(\bfF,\bftau,\bfM)=2J^{-1}\bfF\frac{\partial W}{\partial \bfC} \left(\bfC,\bftau,\bfG\right)\bfF^T.\label{eqn:ct}
\end{equation}
In the same way, the statement of ISRI given in \eqref{eqn:1st} can be reframed as
\begin{equation}
    J^{-1}W\left(\bfF^T\bfF,\bftau,\bfG\right)=\tilde{J}^{-1}W\left(\tilde{\bfF}^T\tilde{\bfF},\hat{\bftau},\hat{\bfM}\otimes\hat{\bfM}.\right),
    \label{eqn:2nd1}
\end{equation}
or
\begin{equation}
    W\left(\bfF^T\bfF,\bftau,\bfG\right)=\tilde{J}W\left(\tilde{\bfF}^T\tilde{\bfF},\bfsigma(\hat{\bfF},\bftau,\bfM),\hat{\bfF}\bfG\hat{\bfF}^T\right).\label{eqn:2nd}
\end{equation}
This relation determines an important restriction for initially stressed materials.

To see this, we consider $\hat{\bfF}$ as any proper rotation tensor $\bfQ$ and determine the relationships between $\bftau$ and $\hat{\bftau}$, and $\bfM$ and $\hat{\bfM}$. Note that if we additionally consider $\tilde{\bfF}=\bfI$, we obtain the deformation gradient $\bfF=\hat{\bfF}=\bfQ$, and consequently $\bfC={\bfF}^T{\bfF}=\bfI$.
Then, because  $\tilde{\bfF}$ is chosen to be the identity, the current configuration $\mathcal{C}$ is same as the configuration $\hat{\mathcal{R}}$ and the Cauchy stress in the configuration $\mathcal{C}$ is identical to the initial stress $\hat{\bftau}$ in configuration $\hat{\mathcal{R}}$.
We obtain this initial stress $\hat{\bftau}$ from \eqref{eqn:ct} as
\begin{equation}
    \hat{\bftau}=\bfsigma(\hat{\bfF},\bftau,\bfM)\big|_{{\hat{\bfF}}=\bfQ}=2\bfQ\left[\frac{\partial W}{\partial \bfC} \left(\bfC,\bftau,\bfG\right)\right]_{\bfC=\bfI}\bfQ^T.\label{tau1}
\end{equation}
However, using initial condition $\bfF=\bfI$ in \eqref{eqn:ct}, we obtain the initial stress $\bftau$ in configuration $\mathcal{R}$ as
\begin{equation}
    \bftau = 2\left[\frac{\partial W}{\partial \bfC} \left(\bfC,\bftau,\bfG\right)\right]_{\bfC=\bfI}.
    \label{tau}
\end{equation}
Substituting \eqref{tau} in \eqref{tau1}, we obtain $\hat{\bftau}=\bfQ\bftau\bfQ^T$.
Moreover, substituting $\hat{\bfF}=\bfQ$ in \eqref{eqn:texture-reln} we obtain $\hat{\bfM}=\bfQ\bfM$.
Employing the above expressions of $\hat{\bftau}$ and $\hat{\bfM}$ in \eqref{eqn:2nd1} we find the following important restriction on the strain energy density,
\begin{equation}
    W\left(\bfC,\bftau,\bfG\right)=W\left(\bfQ\bfC\bfQ^T,\bfQ{\bftau}\bfQ^T,\bfQ{\bfG}\bfQ^T\right).\label{eqn:ISRI}
\end{equation}

The 19 invariants which satisfy \eqref{eqn:ISRI} can be chosen as follows,
\begin{align}
   & I_1= \trace\bfC,
   && I_2= \tfrac{1}{2}[\trace(\bfC)^2 - \trace(\bfC^2)],
   && I_3=\text{det}\;\bfC,
   && I_{\bfM} = \bfM. \bfM,
   \nonumber
   \\[4pt]
   &  I_{4} = \bfM.\bfC\bfM,
   &&   I_5 = \bfM.\bls{C}^2\bfM,
   && I_{\bftau_1}=\trace\bftau,
   &&
   \nonumber
   \\[4pt]
   & I_{\bftau_2}= \tfrac{1}{2}[\trace(\bftau)^2 - \trace(\bftau^2)],
   &&  I_{\bftau_3}=\text{det}\left(\bftau\right),
   &&  I_{\bftau_4}=\bfM.\bftau\bfM,
   && I_{\bftau_5} = \bfM.\bftau^2\bfM,
   \nonumber
   \\[4pt]
   & I_9= \trace(\bls{\tau}\bls{C}),
   &&  I_{10} = \trace( \bls{\tau}\bls{C}^2),
   && I_{11} = \trace(\bls{\tau}^2 \bls{C}),
   && I_{12} = \trace(\bls{\tau}^2 \bls{C}^2),
   \nonumber
   \\[4pt]
   & I_{13}= \bls{M}.\bls{\tau}\bls{C}\bls{M},
   && I_{14}= \bls{M}.\bls{\tau}\bls{C}^2\bfM,
   && I_{15}= \bls{M}.\bls{\tau}^2\bls{C}\bfM,
   && I_{16}= \bls{M}.\bls{\tau}^2\bls{C}^2\bfM.
   \label{inv4}
\end{align}
Note that we allow $\bfM \cdot \bfM \not = 1$ (hence the presence of $I_\bfM$ in the list), as it simplifies how we use equation~\eqref{eqn:2nd1} to restrict $W$
(Alternatively, we could have imposed $\bfM \cdot \bfM = 1$, which would complicate calculations later, but would still lead to the same results.)

For background on the invariant approach to constitutive equations, see Spencer \cite{spencer1971part} and Zheng \cite{zheng1994theory}.
Mukherjee and Mandal \cite{mukherjee2021generalized} used similar invariants for a generalisation using fractional powers of $\bfC$, and Ogden and Singh \cite{ogden2011propagation} used some of these invariants to study wave propagation in initially stressed solids.
Shariff \textit{et al.} \cite{shariff2017spectral}  show, using a spectral decomposition, that each standard invariant used for isotropic materials with initial stress is not independent of the others, making it is likely that some of the invariants in \eqref{inv4} can be expressed in terms of the others.

Using \eqref{eqn:ct}, we next derive the Cauchy stress from the above-derived invariants as
\edit{\begin{align}
    \bfsigma&\left(\bfF,\bfM,\bftau\right)=2J^{-1}\left[W_{I_1}\bfB+W_{I_2}\left(I_1\bfB-\bfB^2\right)+I_3W_{I_3}\bfI+W_{I_4}\bfF\bfG\bfF^T+W_{I_{11}}\left(\bfF\bftau^2\bfF^T\right)\right]\nonumber\\
  &+2J^{-1}\left[W_{I_5}\left(\bfF\bfG\bfF^T\bfB+\bfB\bfF\bfG\bfF^T\right)+W_{I_{9}}\left(\bfF\bftau\bfF^T\right)+W_{I_{10}}\left(\bfF\bftau\bfF^T\bfB+\bfB\bfF\bftau\bfF^T\right)\right]\nonumber\\
     &+2J^{-1}\left[W_{I_{12}}\left(\bfF\bftau^2\bfF^T\bfB+\bfB\bfF\bftau^2\bfF^T\right)+W_{I_{13}}\left(\bfF\bftau\bfG\bfF^T+\bfF\bfG\bftau\bfF^T\right)+W_{I_{15}}\left(\bfF\bfG\bftau^2\bfF^T\right)\right]\nonumber\\
    &+J^{-1}\left[W_{I_{14}}\left(\bfF\bfG\bftau\bfF^T\bfB+\bfB\bfF\bftau\bfG\bfF^T+\bfB\bfF\bfG\bftau\bfF^T+\bfF\bftau\bfG\bfF^T\bfB\right)+W_{I_{15}}\left(\bfF\bftau^2\bfG\bfF^T\right)\right]\nonumber\\
    &+J^{-1}\left[W_{I_{16}}\left(\bfF\bfG\bftau^2\bfF^T\bfB+\bfB\bfF\bftau^2\bfG\bfF^T+\bfB\bfF\bfG\bftau^2\bfF^T+\bfF\bftau^2\bfG\bfF^T\bfB\right)\right],\
  \label{eqn:cauchy}
\end{align}}
where $W_{I_i} :=\partial W/\partial I_i$.
This Cauchy stress should satisfy the initial condition $\bftau=\bfsigma(\bfI,\bftau,\bfM)$ for $\bfF=\bfI$, which provides the condition for initial stress compatibility on the material parameters.

Note that the initial condition is $\bftau=\bfsigma(\bfI,\bftau,\bfM)$. Hence by evaluating \eqref{eqn:cauchy} for $\bfF=\bfI$, we write the initial stress compatibility condition as
\begin{align}
    \bftau &=\left(2{W}_{0I_1}+4{W}_{0I_2}+2{W}_{0I_3}\right)\bfI+\left(2{W}_{0I_4}+4{W}_{0I_5}\right)\bfG+\left(2{W}_{0I_{9}}+4{W}_{0I_{10}}\right)\bftau\nonumber\\
    &+\left(2{W}_{0I_{11}}+4{W}_{0I_{12}}\right)\bftau^2+\left(2{W}_{0I_{13}}+4{W}_{0I_{14}}\right)\left(\bfG\bftau+\bftau\bfG\right)\nonumber\\
   &+\left(2{W}_{0I_{15}}+4{W}_{0I_{16}}\right)\left(\bftau^2\bfG+\bfG\bftau^2\right),\label{eqn:compatibility}
\end{align}
Equating coefficients of $\bfI$, $\bftau$, etc., on both sides of \eqref{eqn:compatibility} we have
\begin{align}
  &\left(2{W}_{0I_1}+4{W}_{0I_2}+2{W}_{0I_3}\right)=0,
  && \left(2{W}_{0I_4}+4{W}_{0I_5}\right)=0,
  && \left(2{W}_{0I_{9}}+4{W}_{0I_{10}}\right)=1,
  \notag\\
  &\left(2{W}_{0I_{13}}+4{W}_{0I_{14}}\right)=0,
  && \left(2{W}_{0I_{13}}+4{W}_{0I_{14}}\right)=0,
  && \left(2{W}_{0I_{15}}+4{W}_{0I_{16}}\right)=0
\end{align}
where ${W}_{0I_i}$ is the value of $W_{I_i}$ calculated in the reference configuration for $\bfF=\bfI$.



\section{Strain energy functions satisfying ISRI and application}
\label{sec:example}


In the previous section, we derived Initial Stress Reference Independence (ISRI) for materials with texture-induced and stress-induced anisotropy.
From  now on, we assume that strain energy functions satisfy ISRI, to avoid some non-physical behaviours \cite{gower2015initial,gower2017new,agosti2018constitutive}.

In this section we propose two strain energy densities for initially strained (Section \ref{strained}) and  initially stressed (Section \ref{stressed}) compressible structurally anisotropic materials, one which is linear in $I_9$ and another with nonlinear combinations of the invariants.
Our approach to deriving these strain energy densities can be applied to many other invariant-based energy functions.
Then, as \edit{two important applications, we solve the boundary value problems of an inflated tube (Section \ref{bvpp}) and of the tension of a welded steel plate (Section \ref{wp}).}


\subsection{Constitutive models for initial strain and texture-induced anisotropy}
\label{strained}


One way to deduce strain energies that depend on the stress explicitly and satisfy ISRI, is to start with a conventional strain energy that depends only on the strain.
To this end, we introduce an initial deformation gradient $\bfF_0$ from a stress-free configuration $\mathcal{R}_0$ to the initially strained configuration $\mathcal{R}$, so that the total deformation gradient from $\mathcal{R}_0$ to the deformed configuration $\mathcal{C}$ is  $\bar{\bfF}=\bfF\bfF_0$.
We call $\bfM_0$ the unit vector along the direction of transverse anisotropy in $\mathcal R_0$.

A strain energy density defined as $W:=J_0^{-1}W_0\left({\bar{\bfC}},{\bfM}_0\otimes{\bfM}_0 \right)$ can be expressed as $W=J^{-1}W_1\left(\bfC,{\bfB}_0,{\bfM\otimes\bfM}\right)$ using transformation of reference configuration $\mathcal{R}_0\rightarrow \mathcal{R}$ where $\bar{\bfC}= {\bar{\bfF}}^T{\bar{\bfF}}$, ${\bfB}_0={\bfF}_0{\bfF}_0^T$, and ${\bfM}_0$ is along the preferred direction of structural anisotropy in configurations $\mathcal{R}_0$.



The strain energy density with reference $\mathcal{R}_0$ should be a function $W = W\left(\bar{I}_1,\bar{I}_2, \bar{I}_3, \bar{I}_4, \bar{I}_5, \ldots\right)$, where $\bar{I}_1=\trace{\bar{\bfC}}$, $\bar{I}_2=[\trace\left(\bar{\bfC}\right)^2 - \trace\left(\bar{\bfC}^2\right)]/2$, $\bar{I}_3=\mathrm{det}\;{\bar{\bfC}}$, $\bar{I}_4=\bfM_0.{\bar{\bfC}}\bfM_0$, $\bar{I}_5=\bfM_0.{\bar{\bfC}}^2\bfM_0$, etc., are various invariants of ${\bar{\bfC}}$.
When $\mathcal{R}$ is considered as the reference, we use ${\bar{\bfF}}=\bfF\bfF_0$ to rewrite the first five invariants of ${\bar{\bfC}}$ as
\begin{align}
   \bar{I}_1& =\trace\left(\bfB_0\bfC\right),\label{I1}\\
     \bar{I}_2&= \tfrac{1}{2}\left\{\left[\trace\left(\bfB_0{\bfC}\right)\right]^2 - \trace\left[\left(\bfB_0{\bfC}\right)^2\right]\right\},\label{I_2}\\
     \bar{I}_3&=J_0^2J^2,\label{I_3}\\
     \bar{I}_4&=\bfM.\bfC\bfM,\label{I_4}\\
          \bar{I}_5&=\bfM.\bfC\bfB_0\bfC\bfM,\label{I_5}
\end{align}
where $J_o=\text{det} \;\bfF_0$ and $\bfM=\bfF_0\bfM_0$.
Notice that the additional structural tensors introduced in the above invariants are $\bfB_0$ and $\bfF_0\bfM_0\otimes\bfM_0\bfF_0^T$.
Using the invariants \eqref{I1}-\eqref{I_5}, we propose the following two strain energy densities for initially strained compressible transversely isotropic materials
\begin{align}
   W&= \frac{\mu_1}{2J_0}\left(\bar{I}_1-3\right) + \frac{\mu_2}{2J_0} \left( \bar{I}_4-1\right) - \frac{\mu_1}{ J_0 \beta}
   \left[1-\left(\sqrt{\bar{I}_3}\right)^{-\beta}\right],
   \label{w-1}
   \\[4pt]
    W&= \frac{\mu_1}{2J_0}\left(\bar{I}_1-3\right) + \frac{\mu_2}{2J_0} \left(\bar{I}_5-1\right)- \frac{\mu_1}{ J_0 \beta}
   \left[1-\left(\sqrt{\bar{I}_3}\right)^{-\beta}\right],
\label{w-2}
\end{align}
for the stress-free reference configuration, where $\mu_1$, $\mu_2$ and $\beta$ are material parameters.
The volumetric function in \eqref{w-1} and \eqref{w-2} is chosen by following Haughton and Orr \cite{haughton1997eversion}, but other choices are available in the literature.
This function requires $\beta>-{1}/{3}$ to ensure a positive initial bulk modulus.

The Cauchy stress $\bfsigma=\frac{2}{J}\bfF\frac{\partial W}{\partial \bfC}\bfF^T$ is then determined for the strain energy functions \eqref{w-1} and \eqref{w-2} as
\begin{align}
   \bfsigma&= \frac{\mu_1}{JJ_0}\bfF\bfB_0\bfF^T + \frac{\mu_2}{JJ_0} \bfF\bfM\otimes\bfF\bfM - \frac{\mu_1} {\left(JJ_0\right)^{\beta+1}}\bfI,
   \label{s1}
   \\
    \bfsigma&= \frac{\mu_1}{JJ_0}\bfF\bfB_0\bfF^T+\frac{\mu_2}{JJ_0} \left(\bfF\bfB_0\bfC\bfM\otimes\bfF\bfM +\bfF\bfM\otimes\bfM\bfC\bfB_0\bfF^T\right)-\frac{\mu_1} {\left(JJ_0\right)^{\beta+1}}\bfI,
    \label{s2}
\end{align}
respectively.

Gower \textit{et al.} \cite{gower2017new}  show that initially strained models naturally satisfy ISRI, so that it is the case  for the proposed strain energy functions \eqref{w-1}, \eqref{w-2}.
These strain energy functions are useful when the initial strain is known (in terms of $\bfB_0$, $\bfM\otimes\bfM$, etc.) instead of the initial stress $\bftau$.
In the next section, we determine strain energy functions for initially stressed transversely isotropic materials when the initial strain is not known.


\subsection{Constitutive models for initial stress and texture-induced anisotropy}
\label{stressed}


Here, we rewrite  the strain energy densities \eqref{w-1}-\eqref{w-2} using the invariants of initial stress symmetry derived in Section \ref{isrisec}.
We eliminate $\bfF_0$ and related quantities in favour of the initial stress $\bftau$.
The end result is that the final expressions of $W$ do not depend on knowing the stress-free configuration $\mathcal R_0$, which indeed might remain hypothetical and unattainable.

Starting with \eqref{w-1}, we first obtain the expression of the initial stress $\bftau$ by substituting $\bfF=\bfI$ in \eqref{s1}, which gives
\begin{equation}
     \frac{\mu_1}{J_0}\bfB_0+\frac{\mu_2}{J_0} \bfM\otimes\bfM=\bftau+\frac{\mu_1} {J_0^{\beta+1}}\bfI.
     \label{s11}
\end{equation}
Next we multiply this equation by $\bfC$ and take the trace, to find the following expression,
\begin{equation}
    W=\frac{1}{2}\left(I_9-3\mu_1-\mu_2\right) + \frac{\mu_1} {2J_0^{\beta+1}}I_1-\frac{\mu_1}{J_0 \beta}\left[1-\left(JJ_0\right)^{-\beta}\right].
    \label{wl13}
\end{equation}
It remains to eliminate  $J_0$ to obtain a strain-energy function in terms of initial stress only (and not initial strain).
To express $J_0$ in terms of initial stress we take the determinant of both sides of \eqref{s11}, as
\begin{equation}
    \text{det} \left({\mu_1}\bfB_0+{\mu_2} \bfM\otimes\bfM\right) = J_0^3\text{det}\left(\bftau+\frac{\mu_1} {J_0^{\beta+1}}\bfI\right),
    \label{s1-1}
\end{equation}
 the left hand side of which is calculated as
 \begin{equation}
 \text{det}\left({\mu_1}\bfB_0+{\mu_2} \bfF\bfM_0\otimes\bfM_0\bfF^T\right)
 = \left(\text{det}\;\bfF_0\right)\text{det}\left({\mu_1}\bfI+{\mu_2} \bfM_0\otimes\bfM_0\right)\left(\text{det}\bfF_0\right)
 =J_0^2 \mu_1^2\left(\mu_1+\mu_2\right).
  \label{det1}
\end{equation}
Substituting \eqref{det1} in \eqref{s1-1} and expanding the determinant on the right hand side, we obtain an equation for $J_0$,
\begin{equation}
\left(\frac{\mu_1} {{J_0}^{\beta+1}}\right)^3 + \left(\frac{\mu_1} {{J_0}^{\beta+1}}\right)^2 I_{\bftau_1} + \left(\frac{\mu_1} {{J_0}^{\beta+1}}\right) I_{\bftau_2} + I_{\bftau_3}  - \frac{\mu_1^2}{J_0}\left(\mu_1+\mu_2\right) = 0.
\label{mmm}
\end{equation}

For a given initially stressed solid with transverse texture, $\mu_1$, $\mu_2$, $\bftau$ and $\beta$ are prescribed, and $J_0$ is a quantity to found by solving this equation numerically.
Then the strain energy density  \eqref{wl13} is explicit, and gives the Cauchy stress by differentiation.
For example, by choosing $\mu_1=3.0$, $\mu_2=1.5$, $\beta=1.0$, $\bftau=\text{diag}\left(1.12, 1, 1.5\right)$, we find $J_0=1.9157$ as the only real positive root of \eqref{mmm}.

In this way, the model is useful for modelling residually-stressed materials, without having to specify the origin of the residual stress.
The resulting constitutive relation is
\begin{equation}
  \bfsigma = \frac{1}{J} \left[ \bfF\bftau\bfF^T + \frac{\mu_1}{J_0^{\beta+1}} \left(\bfB - \frac{1}{J^{\beta+1}}\bfI\right)\right].
  \label{stress1}
  \end{equation}
Although it is not obvious at first glance, texture anisotropy is indeed embedded into the model \eqref{wl13},\eqref{stress1}. It is measured by the magnitude of the material parameter $\mu_2$, which plays a role in determining $J_0$ from Equation  \eqref{mmm}, and also be seen in the form that the initial stress $\bftau$ must take according to \eqref{s11}.

Turning now to \eqref{w-2}, we find that a similar, but more involved, process (detailed in Appendix \ref{AppendixA}), leads to the following expression for $W$,
\begin{align}
    W= &\frac{1}{2}\left(I_9 + \frac{\mu_1} {{J_0}^{\beta+1}}I_1 - 3\mu_1\right)
    + \frac{\mu_1}{2} \left[2 a_2 \left(I_{13} + \frac{\mu_1} {{J_0}^{\beta+1}}I_4\right) + a_3 \left(I_{\bftau_4} I_4 + \frac{\mu_1} {{J_0}^{\beta+1}}I_4I_{\bfM}\right)\right]
    \nonumber
    \\
    & + \frac{\mu_2}{2} \left[a_1 \bfM.\bfC\bftau\bfC\bfM + 2a_2I_4I_{13} + a_3I_4^2I_{\bftau_4} + \frac{\mu_1} {{J_0}^{\beta+1}}\left(a_1 I_5 + 2 a_2 I_4^2 + a_3 I_4^2 I_{\bfM}\right) - 1\right]
    \nonumber\\
    &-{\mu_1}\frac{(1-\left(JJ_0\right)^{-\beta})}{\beta J_0},\label{ef}
\end{align}
where the material parameters $a_1$, $a_2$, and $a_3$ are calculated as
\begin{equation}
    a_1=\frac{1}{\mu_1},
    \qquad
    a_2 = - \frac{\mu_2}{\mu_1(\mu_1 + \mu_2 I_{\bfM})},
    \qquad
    a_3 = \frac{2 \mu _2^2}{\mu_1 \left(\mu_1 ^2+3 \mu_1  \mu_2 I_{\bfM} + 2 \mu _2^2 I_{\bfM}^2 \right)}
    \label{a123}
\end{equation}
by inverting a fourth order tensor \cite{jog2006derivatives} and the non-linear constitutive relation for transversely isotropic materials. \edit{The expression of initial strain obtained in terms of stress and texture is given by
\begin{equation}
  \bfB_0=a_1\hat{\bftau}+a_2 \hat{\bftau}\bfM\otimes\bfM+ a_2\bfM\otimes\bfM\hat{\bftau}+a_3\bfM\otimes\bfM\hat{\bftau}\bfM\otimes\bfM,\label{invert2}
\end{equation}
with $\hat{\bftau}=J_0\left(\bftau +  \tfrac{\mu_1} {{J_0}^{\beta+1}}\bfI\right)$, which is substituted in \eqref{s2} to obtain Cauchy stress from a stressed reference.
}

The expression \eqref{ef} of $W$ contains most of the invariants developed in \eqref{inv4}: $I_1$, $I_3$, $I_{\bfM}$ $I_4$, $I_5$, $I_{\bftau_4}$, $I_9$, $I_{13}$, and also $\bfM.\bfC\bftau\bfC\bfM$, which can be expressed as a function of other standard invariants using the Cayley-Hamilton theorem.

\edit{Using this constitutive relation requires a few more calculations, such as computing the initial strain invariants like $J_0$ and $\bfM_0.\bfC_0^2\bfM_0$ in terms of initial stress invariants. We show these details in Appendix \ref{AppendixA}.}
%



\subsection{Inflation of a \edit{stressed transversely isotropic compressible} tube}
\label{bvpp}


Here we solve the boundary value problem of the inflation of a thick-walled cylindrical tube with  initial stress and preferred direction of texture anisotropy aligned with (a) the axis of the cylinder and (b) the radial direction. \edit{This modelling aims at capturing the residual stress fields observed in tubular biological structures such as arteries, veins, ducts, etc., and also leeks or squid rings as shown in Figure \eqref{fig:1}.}

The deformation gradient for inflating the tube is taken as
\beq
\bfF = \text{diag}\left(\frac{\mathrm{d}r}{\mathrm{d}R},\frac{r}{R},\frac{z}{Z}\right),
\eeq
where $r$ and $R$ are the radii in the current and the reference configurations respectively, $z$ and $Z$ are the axial positions of a material point in the wall.
We assume that the tube is constrained so that no axial extension or contraction occurs: $z=Z$, for example by being placed between two fixed rigid platens~\cite{du2019prescribing}.

We call $R_A$ and $R_B$ the inner and outer radii of the tube in the reference configuration.
For the radial component of the initial stress we choose $ \tau_{RR} = \Lambda\left(R-R_A\right)\left(R-R_B\right)$ where $\Lambda$ is a measure of the initial stress magnitude, which may be positive (radially tensile) or negative (compressive).
This choice leaves the curved faces free of normal stress (cylinder in a vacuum).
The circumferential component is found by solving the equation of equilibrium $\mathrm{d}\tau_{RR}/\mathrm{d}R + (\tau_{RR}-\tau_{\theta\theta})/R = 0$, to give $\tau_{\Theta\Theta}=\Lambda \left[\left(R-R_A\right)\left(R-R_B\right) - R \left(R_A+R_B-2R\right)\right]$.
Finally $\tau_{ZZ}$ is found from the boundary conditions at the platens.
\begin{figure}[!htbp]
    \centering
    \includegraphics[width=0.8\textwidth]{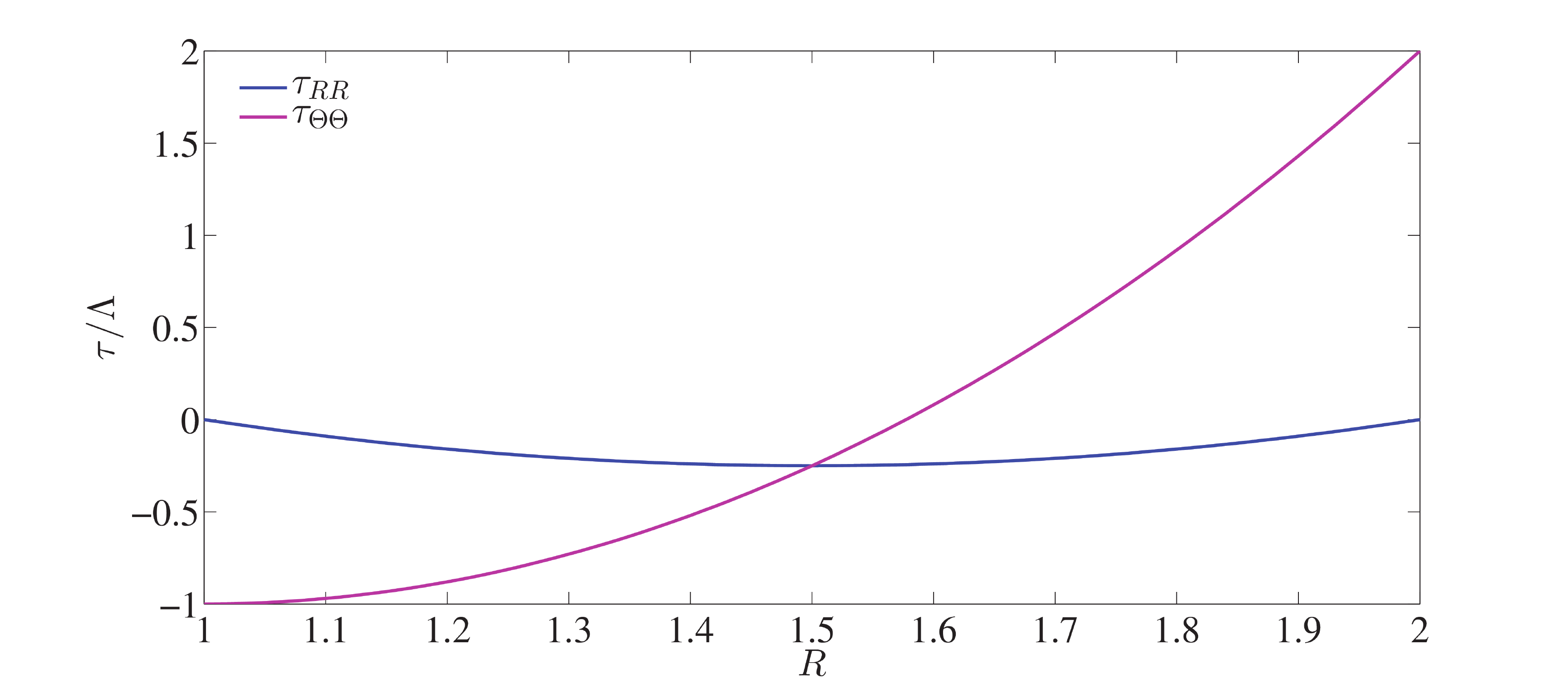}
    \caption{\edit{Radial distribution of the residual stress field in the initially stressed tube. The radial component vanishes on the curved faces of the tube to satisfy the traction-free surface conditions. The hoop stress is compressive on the inner circumference and tensile on the outside. This field captures the characteristics of the residual stress observed in real-world structures such as arteries, squid rings, leeks, etc., which open up when cut radially (see Figure \ref{fig:1}). The parameter $\Lambda$ determines the magnitude of the residual stress.}}
    \label{residual_stress}
\end{figure}
\edit{Figure \ref{residual_stress} shows the spatial distribution of the residual stress field.
It agrees with the residual stress fields reported in Figure 4 of \cite{raghavan2004three} for arteries, and can also be used for other tubular structures which open up when cut radially.
Other distributions can be used to model this opening (logarithmic, exponential, etc.), but as shown by Ciarletta \textit{et al.}~\cite{ciarletta2016morphology}, we can expect the qualitative results to be similar.
}

The stress components in this inflated cylinder corresponding to the strain energy density \eqref{wl13} are
\begin{equation}
\sigma_{rr} = \left(\tau_{RR} + \frac{\mu_1}{J_0^{\beta+1}}\right) \frac{r'R}{r} - \frac{\mu_1}{(JJ_0)^{\beta+1}}, \qquad
\sigma_{\theta\theta} = \left(\tau_{\Theta\Theta} + \frac{\mu_1}{J_0^{\beta+1}}\right) \frac{rr'}{R}- \frac{\mu_1}{(JJ_0)^{\beta+1}},
\label{stt}
    \end{equation}
where $r'={\mathrm{d}r}/{\mathrm{d}R}$ and $J=r'r/R$.
To find the  function $r = r(R)$, we write the equilibrium equation $ \mathrm{d}\sigma_{rr}/\mathrm{d}r + (\sigma_{rr}-\sigma_{\theta\theta})/r=0$ as
\begin{equation}
    \frac{\mathrm{d}\sigma_{rr}}{\mathrm{d}R}+\frac{r'\left(\sigma_{rr}-\sigma_{\theta\theta}\right)}{r}=0.
    \label{eqn}
\end{equation}
For our example, we assume that a hydrostatic pressure $-P$ (to be determined) is applied at the inner face $r(R_A)$, such that the inner diameter increases by 50\%.
Hence the boundary conditions are
\begin{equation}
    r(R_A) = 1.50R_A, \qquad\sigma_{rr}\left(R_B\right)=0.
    \label{bcs}
\end{equation}

\begin{figure}[h!]
    \centering
    \includegraphics[width=0.9\textwidth]{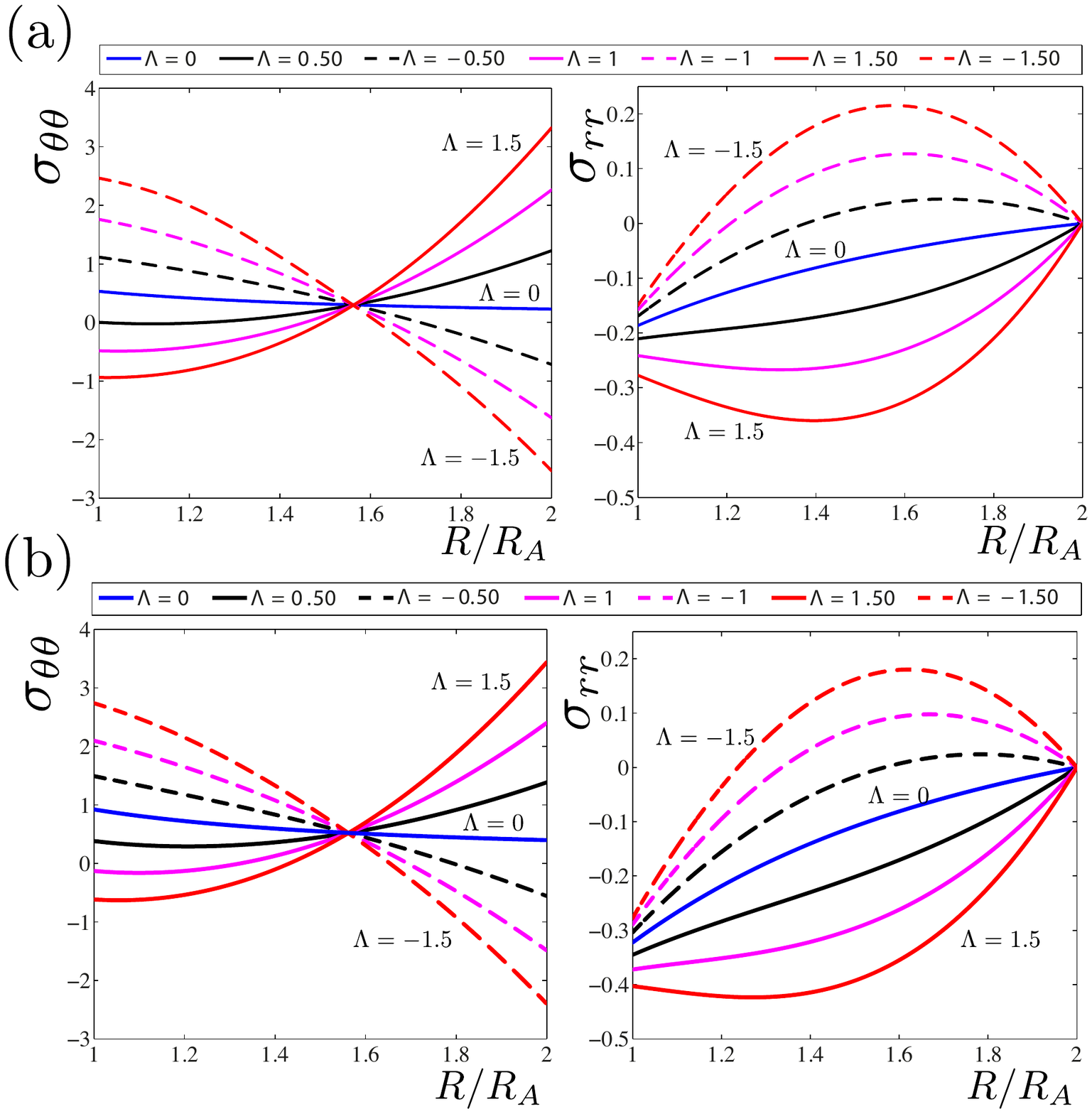}
    \caption{Stress distribution for various amounts of residual stress as measured by $\Lambda = -1.5, -1.0, 0.0, 1.0, 1.5$, for a compressible tube with texture anisotropy modelled by the strain-energy density \eqref{wl13}.
    Here the importance of anisotropy to initial stress is measured by the ratio $\mu_2/\mu_1=1.5$.
    We consider that under the pressure $-P = \sigma_{rr}(R_A)$, the inner radius has increased to $r(R_A)=1.5 R_A$, while the outer face at $R_B = 2.0$ remains free of traction.
    Circumferential and radial stress components when the texture direction is aligned with (a) the cylinder axis and (b) the radial direction.}
    \label{fig:my_label}
\end{figure}
To solve numerically  \eqref{eqn}, subject to \eqref{bcs}, we use the MATLAB `\emph{bvp5c}' command.
Figure \ref{fig:my_label} shows the variations of the Cauchy stress components $\sigma_{RR}$ and $\sigma_{\theta\theta}$ through the thickness for different amounts of the initial stress parameter $\Lambda$.
The internal pressure required to inflate the cylinder is found as $P= -\sigma_{RR}\left(R_A\right)$.

\edit{We observe that all the hoop stress $\sigma_{\theta\theta}$ curves cross at a point through the thickness, independently of the magnitude of the residual stress.
A similar scenario was also observed in the investigation of unbending of an incompressible stressed \emph{isotropic} cylinder \cite{mukherjee2021generalized}.
Using a closed-form solution, Mukherjee and Mandal \cite{mukherjee2021generalized} showed that the hoop and radial components of residual stress have opposing effects and nullify each other's contribution at this particular point. Such a closed-form solution is not attainable for the present problem of cylinder inflation where the material is compressible and includes texture symmetry as well.
}

\edit{For positive or negative $\Lambda$, $\sigma_{rr}$ is distributed \emph{almost} symmetrically around the $\Lambda = 0$ curve (no residual stress).
It would be exactly symmetric if we considered that the total stress is the sum of the residual stress and of the stress developed without residual stress (see residual stress distribution in Figure \ref{residual_stress}). However, due to the intrinsic nonlinearity of the system, exact symmetry is not observed.
}

In conclusion, we observe that the residual stress and the texture anisotropy significantly affect both stress distribution characteristics for this boundary value problem.


\subsection{\edit{Tension of a welded steel plate}}
\label{wp}


\edit{
Residual stress develops during manufacturing due to accumulation of incompatible plastic strain. In this section, we investigate the stress developed during the stretching of a welded steel plate by using our strain energy functions \eqref{w-2} and \eqref{w-1}.

Figure \ref{fig:1} (d) shows the residual stress in a welded alloy steel structure, which we obtained through finite element analysis.
There, the weld plate has finite dimensions, and because of end effects, the residual stress field is complicated, as seen in the figure.

For an infinitely long weld joint, Masubuchi and Martin \cite{masubuchi1966investigation,choudhury2020temperature}  determined analytically the residual stress field as
\begin{equation} \label{eqn:weld-residual-stress}
    \tau_{yy}\left(x\right)=\tau_0\left[1-\left(\frac{y}{b}\right)^2\right]e^{\left(-\frac{y^2}{2b^2}\right)},
\end{equation}
where $y$ is the direction of welding, $x$ is the transverse direction, $z$ is the direction through thickness, $\tau_0$ is the maximum residual stress, which can be as high as the yield stress of the material, and $b$ is half the thickness of the weld-zone.
We consider some typical values, say $\tau_0=100$ MPa and $b=5$ cm, and report in Figure \ref{f33} the corresponding residual stress distribution.

Using the residual stress~\eqref{eqn:weld-residual-stress}, we can now predict the stress in the weld when under tension.
Note that the considered strain energy densities \eqref{w-1} and \eqref{w-2} are consistent with the Hooke law of linearised elasticity in the stress-free reference for small strain.
For example, when we take $\mu_1=E/(1+\nu)=156.28$ GPa, $\beta=\nu/\left(1-2\nu\right)=0.6363$, $\mu_2=0$, the models match exactly the isotropic properties of steel at small strain where the Young modulus is $E=200$ GPa and the Poisson ratio is $\nu = 0.28$.
To demonstrate the flexibility of our model, we further consider texture anisotropy by taking $\mu_2=0.1\mu_1$, say, and have the texture aligned with the $x$ direction.

We consider that a stress $\sigma_{xx}=90$ MPa is applied in the transverse direction while $\sigma_{zz}=0$.
We solve these equations numerically at every point to obtain $\lambda_x$ and $\lambda_z$, considering that the plane strain condition $\lambda_y=1$ holds, which is reasonable as the plate is infinitely long in the $y$ direction.
The resulting stretch components $\lambda_x$ and $\lambda_y$ are substituted in the current stress \eqref{s1} to find
\begin{equation}
    \sigma_{yy}=\frac{1}{\lambda_x\lambda_z}\left[\tau_{yy}+\frac{\mu_1}{J_0^{\beta+1}}\left(1-\frac{1}{\lambda_x^{\beta+2}\lambda_z^{\beta+2}}\right)\right],
\end{equation}
see the resulting the distribution of $\sigma_{yy}$ in Figure \ref{f33}.
\begin{figure}
    \centering
    \includegraphics[width=0.85\textwidth]{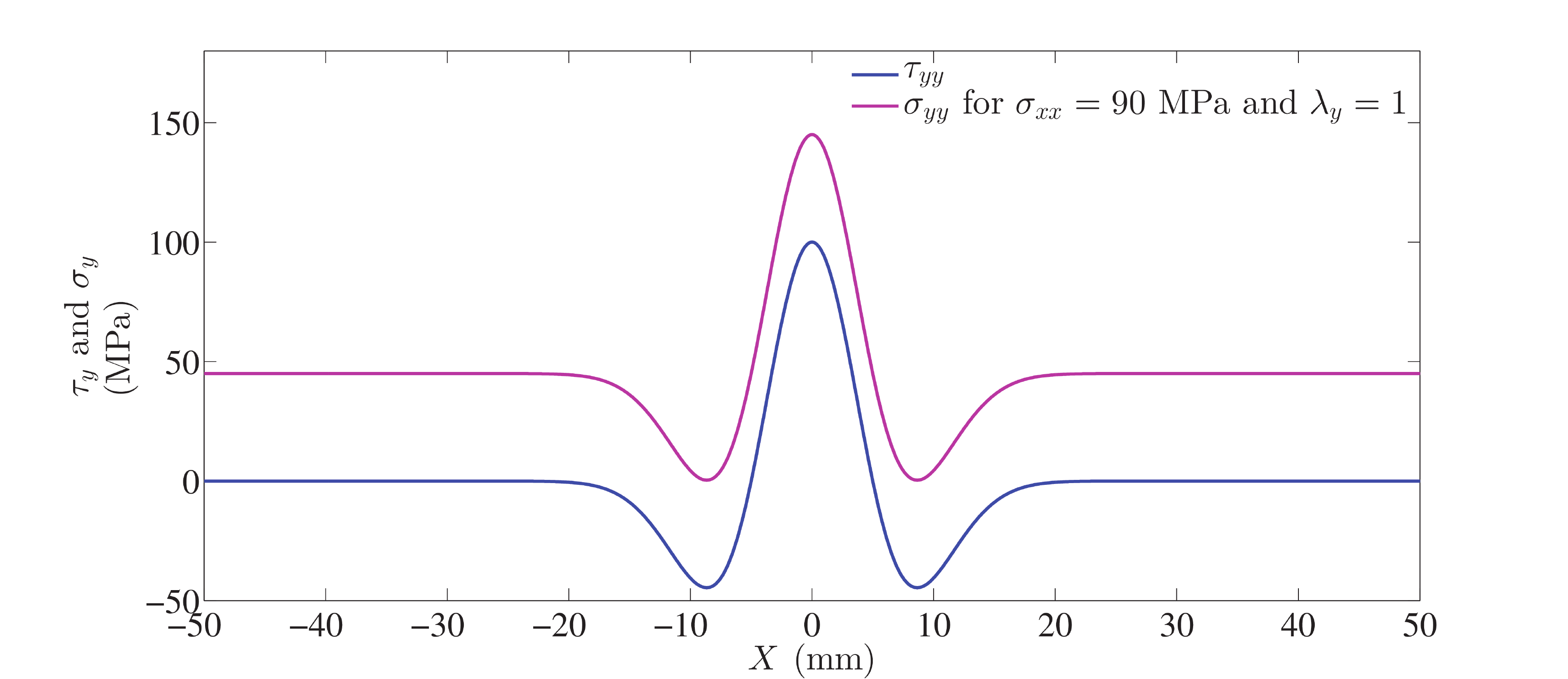}
    \caption{Bottom curve: distribution of the residual stress $\tau_{yy}$ across the width of a welded plate. Top curve: distribution of the Cauchy stress $\sigma_{yy}$ when a stress $\sigma_{xx}=90$ MPa is applied in the transverse direction.}
    \label{f33}
\end{figure}

These results highlight the influence of residual stress on the stress distribution in a welded structure under tension. They show how the high stress intensity in the weld-zone is due to the residual stress, and that the welded structures are expected to fail in the weld zone when stretched from the two sides.
Hence, the models and framework developed in this paper can be used to solve complex engineering problems by providing benchmark solutions.
}


\section{Linearised constitutive relations}
\label{Lin_}


Many residually stressed solids with texture anisotropy, such as the typical metals used in mechanical and civil engineering, can only sustain small elastic deformations.
A linearised model of initial stress is then often required in their modelling.
Here we turn our attention to linearised theories for small strain and small rotation (Section \ref{sec:linearisation}), and then for small initial stress (Section \ref{sec:third-order}).
\edit{The resulting equations allow us to study the propagation of small-amplitude elastic shear waves, with a view to perform the non-destructive evaluation of the level of initial stress (Section \ref{sec:shearwaves}).}
In Section \ref{sec:lin}, we develop a linearised set of equations for ISRI, and further find restrictions that should apply to the material parameters for small initial stress in Section~\ref{sec:small-stress}.
For the calculations, we rely on and make use of results on linearisation from References  \cite{hoger1994positive, gower2017new,destrade2010third,destrade2010third1}.


\subsection{Linearisation for small strain and small rotation}
\label{sec:linearisation}


For small deforations, we write the deformation gradient in the form $\bfF = \bfI + \nabla \bfu$, where $\bfu$ is the infinitesimal displacement vector.
For small-amplitude motions, the equation of motion is \cite{destrade2013stress}:
\begin{equation} \label{eqn:linear-motion}
\rho \ddot u_k	= \mathcal A_{klmn} u_{m,nl},
\end{equation}
where $\rho$ is the mass density and the elastic moduli $\mathcal A_{klmn}$ can be calculated from the equation
\begin{equation} \label{eqn:initial-stress-moduli}
	\boldsymbol{\mathcal A}:\nabla \bfu = \frac{\partial \bfsigma}{\partial \bfF}\bigg|_{\bfF=\bfI}:\nabla \bfu - \bftau \nabla \bfu^T + \bftau \trace \nabla \bfu,
\end{equation}
where we defined the notations
\begin{equation}
    \left(\frac{\partial\bfA}{\partial\bfD}\right)_{ijkl}=\frac{\partial A_{ij}}{\partial D_{kl}},\qquad\text{ and }\qquad \left(\boldsymbol{\mathcal{C}}:\bfA\right)_{ij} = {\mathcal{C}}_{ijkl}A_{lk},
\end{equation}
for any second-order tensors $\bfA$, $\bfD$ and  fourth-order tensor $\boldsymbol{\mathcal{C}}$.

Note that simply calculating the divergence $\mathrm{div}\,(\boldsymbol{\mathcal A}:\nabla \bfu)$, when $\boldsymbol{\mathcal A}$ is constant in the positions $x_1$, $x_2$, and $x_3$,  leads to the right-hand side of \eqref{eqn:linear-motion}.

As the term $\frac{\partial \bfsigma}{\partial \bfF}:\nabla \bfu$ is essential to solve the equation of motion~\eqref{eqn:linear-motion}, it is helpful to have simple representations for it.
First we split the displacement gradient into
\begin{equation}
    \nabla \bfu=\bfomega+\bfepsilon,
\end{equation}
where $\bfomega = (\nabla \bfu/ - \nabla \bfu^T)/2 $ (antisymmetric) and $\bfepsilon = (\nabla \bfu + \nabla \bfu^T)/2 $ (symmetric) are the infinitesimal rotation and strain, respectively.


From \cite[Equation (4.5)]{gower2017new} we have
\begin{equation}
   \frac{\partial \bfsigma}{\partial \bfF}\bigg|_{\bfF=\bfI}:\bfomega = \bfomega\bftau-\bftau \bfomega.
   \label{eqn:variation}
\end{equation}

The difficult part is to determine $\frac{\partial\bfsigma}{\partial \bfF}|_{\bfF=\bfI}:\bfepsilon$, which we do below. Once this is known we calculate the moduli $\boldsymbol{\mathcal{A}}:\nabla \bfu$ using the above to get
\begin{equation} \label{A_fourth_order}
\boldsymbol{\mathcal{A}}:\nabla \bfu = \frac{\partial \bfsigma}{\partial \bfF}\bigg|_{\bfF=\bfI}:\bfepsilon
+ \nabla \bfu\bftau   + \bftau \trace \bfepsilon
- \bfepsilon\bftau - \bftau  \bfepsilon.
\end{equation}

 To determine $\frac{\partial\bfsigma}{\partial \bfF}|_{\bfF=\bfI}:\bfepsilon$ we could simply differentiate \eqref{eqn:cauchy} with respect to  $\bfF$ (with the help of Mathematica~\cite{Mathematica} or another Computer Algebra Package). Instead we use  a representation theorem for second-order tensors \cite{rivlin1949large,zheng1994theory} for each of the coefficients of the invariants $\trace\bfepsilon$, $\trace\bfepsilon\bftau$, etc.,
 to write $\frac{\partial\bfsigma}{\partial \bfF}|_{\bfF=\bfI}:\bfepsilon$ in the form:
\begin{align}
\frac{\partial\bfsigma}{\partial \bfF}\bigg|_{\bfF=\bfI}:\bfepsilon & = \alpha_1\bfepsilon+\alpha_2\left<\bfG\bfepsilon\right>_s+\alpha_3\left<\bftau\bfepsilon\right>_s+\alpha_4\left<\bftau\bfG\bfepsilon\right>_s+\alpha_5\left<\bfG\bftau\bfepsilon\right>_s\nonumber\\
&+\alpha_6\left<\bftau^2\bfepsilon\right>_s+\alpha_7\left<\bftau^2\bfG\bfepsilon\right>_s+\alpha_8\left<\bfG\bftau^2\bfepsilon\right>_s\nonumber\\
&+\left(\alpha_9\bfI+\alpha_{10}\bftau+\alpha_{11}\bftau^2+\alpha_{12}\bfG+\alpha_{13}\left<\bftau\bfG\right>_s+\alpha_{14}\left<\bftau^2\bfG\right>_s\right)\trace\bfepsilon
\nonumber\\
	&+\left(\alpha_{15}\bfI+\alpha_{16}\bftau+\alpha_{17}\bftau^2+\alpha_{18}\bfG+\alpha_{19}\left<\bftau\bfG\right>_s+\alpha_{20}\left<\bftau^2\bfG\right>_s\right)\trace(\bfepsilon\bftau)
\nonumber\\
	&+\left(\alpha_{21}\bfI+\alpha_{22}\bftau+\alpha_{23}\bftau^2+\alpha_{24}\bfG+\alpha_{25}\left<\bftau\bfG\right>_s+\alpha_{26}\left<\bftau^2\bfG\right>_s\right)\trace(\bfG \bfepsilon)
\nonumber\\
	&+\left(\alpha_{27}\bfI+\alpha_{28}\bftau+\alpha_{29}\bftau^2+\alpha_{30}\bfG+\alpha_{31}\left<\bftau\bfG\right>_s+\alpha_{32}\left<\bftau^2\bfG\right>_s\right) \trace(\bfG\bfepsilon\bftau)
\nonumber\\
&+\left(\alpha_{33}\bfI+\alpha_{34}\bftau+\alpha_{35}\bftau^2+\alpha_{36}\bfG+\alpha_{37}\left<\bftau\bfG\right>_s+\alpha_{38}\left<\bftau^2\bfG\right>_s\right)\trace(\bfepsilon\bftau^2)
\nonumber\\
&+\left(\alpha_{39}\bfI+\alpha_{40}\bftau+\alpha_{41}\bftau^2+\alpha_{42}\bfG+\alpha_{43}\left<\bftau\bfG\right>_s+\alpha_{44}\left<\bftau^2\bfG\right>_s\right) \trace({\bfG \bfepsilon\bftau^2}),
\label{linear1}
\end{align}
where the $\alpha_i$ are scalar functions of the invariants of $\bftau$ and $\bfG = \bfM\otimes\bfM$, and the brackets notation denotes the symmetric part of a tensor: $\langle \bfD \rangle_s = (\bfD + \bfD^T)/2$ for any tensor $\bfD$.
Note that terms such as $\left<\bfG\bfepsilon\bftau\right>_s$, and other higher-order terms in $\bftau$, do not appear as they can be expressed using the terms already present in \eqref{linear1}. 

Some of the coefficients $\alpha_i$ depend on each other due to the symmetry of the elasticity moduli.
Using equations (4.9 -- 4.12) from \cite{gower2017new}, we obtain
\begin{equation}
     \bfA: \frac{\partial \bfsigma}{\partial \bfF}\bigg|_{\bfF=\bfI}:\bfD +\left(\bfA:\bftau\right)\trace\bfD=\bfD: \frac{\partial \bfsigma}{\partial \bfF}\bigg|_{\bfF=\bfI}:\bfA+\left(\bfD:\bftau\right)\trace\bfA,
\end{equation}
for any symmetric $\bfA$ and $\bfD$. By separately substituting $\bfA$ and $\bfD$ for $\bfepsilon$ we find that
\begin{align}
& \alpha_4 =\alpha_5, \quad \alpha_7=\alpha_8\nonumber\\
 &   \alpha_{10} =\alpha_{15}-1,\quad \alpha_{11}=\alpha_{33},\quad \alpha_{12}=\alpha_{21},\quad \alpha_{13}=\alpha_{27},\quad \alpha_{14}=\alpha_{39}\quad\alpha_{17}=\alpha_{34}\nonumber\\
   & \alpha_{18}=\alpha_{22},\quad \alpha_{19}=\alpha_{28},\quad \alpha_{20}=\alpha_{40},\quad \alpha_{36}=\alpha_{23},\quad\alpha_{37}=\alpha_{29},\quad\alpha_{38}=\alpha_{41}\nonumber\\
 &   \alpha_{25}=\alpha_{30}, \quad\alpha_{26}=\alpha_{42}\quad\alpha_{32}=\alpha_{43}. \label{al3}\
\end{align}

Using the above, we can determine the total number of material constants needed to describe several special cases.

The simplest possible case is that of a uniaxial initial stress aligned with the preferred direction of transverse texture anisotropy.
Then, $\bftau = \tau \bfG$, which substituted in \eqref{linear1}, together with (\ref{al3}), leads to
\begin{equation}
   \frac{\partial \bfsigma}{\partial \bfF} (\bfI, \bfG,\bftau) : \bfepsilon   =  b_1 \bfepsilon + b_2\left<\bfepsilon\bfG\right>_s + b_3\bfI\trace\bfepsilon + b_4\left(\bfG\trace\bfepsilon+\bfI \trace(\bfG \bfepsilon )\right)
 + b_5 \bfG \trace(\bfG \bfepsilon ),
 \label{lin_stress-strain1}
\end{equation}
where the $b_k$ are a combination of the $\alpha_{j}$ and $\tau$, with all the $b_k$ being independent when we assume that the $\alpha_{ij}$ and $\tau$ can be chosen independently.
Hence this case requires five different elastic constants to describe the response for any $\bfepsilon$, which is the same number of elastic constants required for the description of transversely isotropic solids in linear elasticity.

Often in both natural and man-made materials the principal directions of initial stress are co-axial with the direction of texture anisotropy. For this case, we have
$\bftau = \text{Diag}( \tau_1,  \tau_2, \tau_3)$,
$\bfG = \text{Diag}(M_1^2,  M_2^2, 0)$.
We may then write $\bftau$ as $\bftau = a\bfI_{2\times 2} + b\bfG + \tau_3\bfe_3\otimes\bfe_3$ where $a = (M_1^2\tau_2-M_1^2\tau_1)/(M_1^2-M_2^2)$, $b=\ (\tau_1-\tau_2)/(M_1^2-M_2^2)$ and substitute that decomposition in \eqref{linear1}, together with (\ref{al3}), to obtain
\begin{multline}
   \frac{\partial \bfsigma}{\partial \bfF} (\bfI, \bfG,\bftau) : \bfepsilon
   = b_1\bfepsilon
	 + b_3\left<\bfepsilon\bfG\right>_s
	 + b_4 \left<\bfe_2\otimes(\bfe_3\cdot \bfepsilon)\right>_s
	 \\
   + b_5 \bfI \trace\bfepsilon
    + b_6 \left[\bfe_3\otimes\bfe_3\trace\bfepsilon+\bfI\left(\bfe_3.\bfepsilon\bfe_3\right)\right]
	 + b_7 \left[\bfG\trace\bfepsilon+\bfI \trace(\bfG \bfepsilon)\right]
	 \\
  + b_8\left[\bfG\left(\bfe_3.\bfepsilon\bfe_3\right) + \bfe_3\otimes\bfe_3  \trace(\bfG \bfepsilon)\right]
  + b_9\bfG \trace(\bfG \bfepsilon),\label{lin_stress-strain1}
\end{multline}
which requires nine elastic constants (same as for orthotropic materials in linear elasticity, often described in terms of three Young's moduli, three Poisson's ratios, and three shear moduli~\cite{anand2020continuum}).

\edit{When $\bftau$ is a general triaxial stress field and $\bfM$ lies on the plane containing two principal directions but does not align with any of them, we find 13 linear elastic constants. By substituting the diagonalised form of initial stress $\bftau=\sum_{i=1}^3{\tau_i\bfe_i\otimes\bfe_i}$ and $\bfM=M_1\bfe_1+M_2\bfe_2$ into \eqref{linear1} and using (\ref{al3}), the components of the Cauchy stress $\bfsigma=\bftau+\tfrac{\partial \bfsigma}{\partial\bfF}\big|_{\bfF=\bfI}:\bfepsilon$, without rotation, in the set of coordinates $\left\{\bfe_1,\bfe_2,\bfe_3\right\}$, are given by
\begin{equation}
    \begin{Bmatrix}
    \sigma_{11}\\
    \sigma_{22}\\
    \sigma_{33}\\
    \sigma_{23}\\
    \sigma_{31}\\
    \sigma_{12}
    \end{Bmatrix}
    =    \begin{Bmatrix}
    \tau_{1}\\
    \tau_{2}\\
    \tau_{3}\\
    0\\0\\0
    \end{Bmatrix}
    + \begin{bmatrix}
    b_{11} & b_{12} & b_{13} & 0 & 0 & b_{16}\\
    b_{12} & b_{22} & b_{23} & 0 & 0 & b_{26}\\
    b_{13} & b_{23} & b_{33} & 0 & 0 & b_{36}\\
    0 & 0 & 0 & b_{44} & b_{45} & 0\\
    0 & 0 & 0 & b_{54} & b_{55} & 0\\
    b_{16} & b_{26} & b_{36} & 0 & 0 & b_{66}
    \end{bmatrix}
        \begin{Bmatrix}
    \epsilon_{11}\\
    \epsilon_{22}\\
    \epsilon_{33}\\
    2\epsilon_{23}\\
    2\epsilon_{31}\\
    2\epsilon_{12}
    \end{Bmatrix}.\label{Nye}
\end{equation}}
Note that linear elastic materials with monoclinic symmetry also require 13 elastic constants \edit{and that they have a constitutive relation \cite{nye1985physical,destrade2001explicit} similar to \eqref{Nye}. Hence, we conclude that when the two principal directions of initial stress and the direction of texture remain in the same plane but do not align with each other, the initially stressed material behaves like a monoclinic material.

When the texture direction does not  lie in the plane of any two principal stress directions, we find triclinic symmetry with $21$ material constants which fully populates the $6\times 6$ stiffness matrix in the constitutive relation.}
\begin{figure}[!htbp]
    \centering
    \includegraphics[scale=1.35]{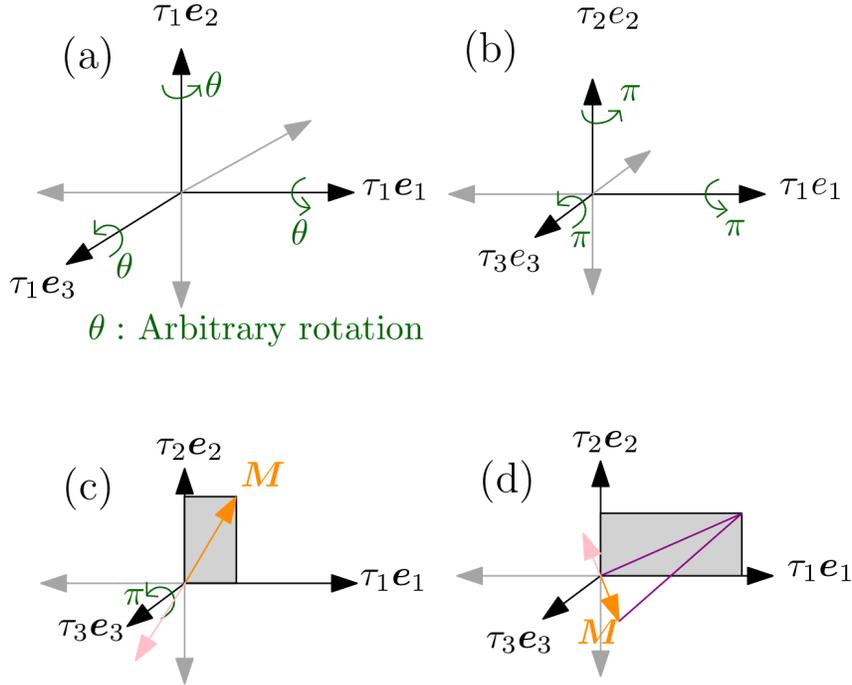}
    \caption{\edit{A visualisation of the symmetries present when combining both residual stress and texture. (a) For hydrostatic initial stress $\left(\tau_1=\tau_2=\tau_3\right)$ without texture, any arbitrary rotation leads to a material with the same properties. We then say that the symmetry group is composed of all rotations, and therefore the symmetry is \emph{isotropy}. (b) For three distinct principal stresses $\left(\tau_1\neq\tau_2\neq\tau_3\neq\tau_1\right)$ without texture, the symmetry group consists of three $180^{\circ}$ rotations about the principal directions $\bfe_1$, $\bfe_2$, and $\bfe_3$ respectively, which is the case of \emph{orthotropy}. (c) When the texture direction is not aligned with any principal direction $\bfe_1$, $\bfe_2$, or $e_3$, but lies on the $\bfe_1-\bfe_2$ plane, {only} a $180^{\circ}$ rotation about $\bfe_3$ axis leaves the material unchanged. This is the symmetry group of \emph{monoclinicity} about the $\bfe_1-\bfe_2$ plane. We obtain 13 linear elastic constants for small deformation theory in this case. (d) Finally, when $\bfM$ is neither aligned with any principal direction, nor lying on any plane passing through any two principal directions, there is no rotation that preserves the structure. This is the case of \emph{triclinicity}, which has 21 linear elastic constants.} }
    \label{fig:nmnm}
\end{figure}

\edit{The above conclusions on material symmetry are in agreement with our investigation of symmetry through physical visualisation in Figure \ref{fig:nmnm}.}
Finally, we note that the $\alpha_{j}$, and therefore $b_j$, may not be independent due to the ISRI restriction~\eqref{eqn:2nd}. We explain this in more detail in Section~\ref{sec:lin}.


\subsection{Third-order elasticity and small initial stress }
\label{sec:third-order}


In the previous section, coefficients such as $\alpha_{j}$ and $b_j$ depend on the combined invariants of $\bftau$ and $\bfG$.
In contrast, here we reduce these scalars to material constants which do not depend on the invariants of $\bftau$ and $\bfG$.
We achieve this reduction by considering that the elastic moduli~\eqref{linear1} vary linearly in $\bftau$.
This is an important case, as it includes the case of third-order weakly nonlinear materials with transverse texture anisotropy and without initial stresss~\cite{destrade2010third}.

The elastic modulus tensor of hyperelastic third-order elastic materials $\boldsymbol{\mathcal A}$ is correct up to first order in $\bfE$, which is asymptotically equivalent to writing $\boldsymbol{\mathcal A}$ as a linear function of $\bftau$. The papers \cite{man1996separation,li2020ultrasonic} show moduli written in terms of the initial stress which are equivalent to the moduli of classical third-order elastic materials.


By dropping all terms which are explicitly of order $\mathcal O(\bftau^2)$ in \eqref{linear1}, and using
 (\ref{al3}), we obtain
\begin{align}
\frac{\partial \bfsigma}{\partial \bfF}\bigg|_{\bfF=\bfI} \hspace{-0.5cm} :\bfepsilon = & \bfepsilon  \left[a_3  (\trace \bftau -\trace(\bfG\bftau))+a_1\right]  +\left< \bfepsilon \bfG \right> \left[a_2-a_3 \trace \bftau\right]
+ a_3 \left< \bftau \bfepsilon \bfG \right> +a_4 \left< \bfepsilon \bftau\right>
    \notag \\
     & + \trace \bfepsilon  \left[\bfI \left(a_5-a_3 (\trace \bftau  - \trace(\bfG\bftau))\right)+a_3 \bfG \trace \bftau
    + a_7 \bftau +a_8 \left<\bftau \bfG \right>+a_6
   \bfG\right]
   \notag \\
   & + \trace(\bfepsilon \bfG) \left[\bfI
   \left(a_3 \trace \bftau +a_6\right)+a_{10} \bftau +a_{11} \left< \bftau \bfG \right> +a_9 \bfG\right]
   \notag \\
	 \label{eqn:stress-linear-tau}
   & + \trace(\bfepsilon \bftau ) \left[\left(a_7+1\right) \bfI+a_{10} \bfG\right] + 2 \trace(\bfepsilon \bfG\bftau ) \left[a_8 \bfI+a_{11} \bfG\right],
\end{align}
%
%
where, for simplicity, we introduced a new set of coefficients $a_j$ which can be related to the $\alpha_j$ in \eqref{linear1}.
Note that there are  11 scalars in total $a_1,a_2, \ldots, a_{11}$. Five of these scalars $a_1, a_2,a_5,a_6$, and $a_9$ depend on the invariants of $\bftau$ and $\bfG$ so they can be expanded in the form
\begin{equation} \label{eqn:alpha-series}
    a_j = a_{j,0} + a_{j,1} \trace \bftau + a_{j,2} \trace (\bftau \bfG),
\end{equation}
which when substituted into $\frac{\partial \bfsigma}{\partial \bfF}:\bfepsilon$ above, clearly shows how many material constants (independent of $\bftau$) are needed to fully characterise the material behaviour:
\begin{align}
&	a_{1,0}, a_{1,1}, a_{1,2}, a_{2,0}, a_{2,1}, a_{2,2}, a_3, a_4, a_{5,0}, a_{5,1}, a_{5,2},
	\notag \\
	\label{linear-as}
&	a_{6,0}, a_{6,1}, a_{6,2},a_7,a_8,a_{9,0}, a_{9,1}, a_{9,2},a_{10},a_{11}.
\end{align}
 Hence, there are 21 material constants, whereas there are only 14 constants needed for the moduli of hyperelastic third-order elastic material with transverse texture anisotropy~\cite{destrade2010third}. How can there be 7 more degrees of freedom between two models which describe the same type of material and to the same asymptotic order? The answer is given in Section~\ref{sec:lin} which shows that the ISRI condition~\eqref{eqn:2nd} leads to exactly 7 equations which connect the constants above.

In polycrystalline materials, such as steel, there are many engineering scenarios where the direction of the texture anisotropy aligns with the direction of stress, and there is a stress-free direction. Examples include steel plates and beams. To represent this scenario we use the identities $\bfG \bftau = \bftau \bfG$, $\trace \bfG = 1$, and the Cayley-Hamilton theorem\footnote{Alternatively, we could choose a coordinate system that diagonalizes $\bftau$ and thus, diagonalises $\bfG$, and then count the number of independent coefficients $\beta_j$. This is best done with a \edit{computer} algebra system~\cite{Mathematica}}
to obtain
\begin{multline} \label{eqn:coaxial-moduli}
\frac{\partial \bfsigma}{\partial \bfF}\bigg|_{\bfF=\bfI} \hspace{-0.5cm} :\bfepsilon = \bfepsilon \bftau + \bftau \bfepsilon -\bftau \trace \bfepsilon +  \beta_1 \bfepsilon +
\beta_2 \bfepsilon  \trace(\bftau \bfG) +
2 \beta_3 \left<\bfepsilon\bfG\right>+
\beta_4 \bfI \trace\bfepsilon \trace(\bftau \bfG) +
\beta _5 \bfG \trace(\bfepsilon \bfG)
\\
+ \beta_6 \left[ \bfI \trace(\bftau \bfG) \trace(\bfepsilon \bfG) + \bftau \bfG \trace\bfepsilon \right] +
2 \beta_7 \trace(\bftau \bfG) \left<\bfepsilon \bfG\right>  +
\beta_8 \bfI \trace\bfepsilon
\\
+ \beta_9 \left[ \bfI \trace(\bfepsilon \bfG) + \bfG \trace\bfepsilon \right] +
\beta_{10} \bftau \bfG \trace(\bfepsilon \bfG) + \bfH,
\end{multline}
where
\begin{multline}
	\bfH = \beta_{11} \bfI \trace \bfepsilon \left[\trace\bftau - \trace(\bftau \bfG)\right] +
	\beta _{12} \trace(\bfepsilon \bfG) \bfI \left[ \trace \bftau -\trace(\bftau \bfG)\right]  +
	\\
	\beta _{12} \left[\bfG \trace \bftau \trace\bfepsilon -\bftau \bfG \trace\bfepsilon \right]+
	\beta_{13} \bfI \left[ \trace(\bfepsilon \bftau) - \trace(\bftau \bfG) \trace(\bfepsilon \bfG)\right]  +
	\beta_{13} \trace\bfepsilon \left[\bftau -\bftau \bfG\right]  +
	\\
	2\beta_{14} \left[\left<\bfepsilon \bftau \right> - \left<\bfepsilon \bfG\right> \trace(\bftau \bfG) \right] +
	\beta_{15} \bfepsilon  \left[\trace \bftau - \trace(\bftau \bfG)\right] +
	 \beta _{16}\trace(\bfepsilon \bfG)  \left[\bfG \trace \bftau - \bftau \bfG\right]  +
	 \\
	 \beta_{17} \left[(\bftau -2 \bftau \bfG) \trace(\bfepsilon \bfG) + \bfG \trace(\bfepsilon \bftau) \right] +
	 2 \beta_{18} \left<\bfepsilon \bfG\right> \left[\trace\bftau - \trace(\bftau \bfG)\right].
\end{multline}
For a uniaxial stress aligned with the texture we would have $\bftau \bfG = \bftau$, $\bfG \trace \bftau = \bftau$, which substituted above leads to $\bfH = \boldsymbol 0$.
Also, similar to the $a_{i,j}$ coefficients, the coefficients $\beta_j$ are not all independent, as we demonstrate in Section \ref{sec:ISRI-aligned-small-stress} below.

Finally, another common simplification is to assume that the coupling between $\bfG$ and $\bftau$ is weak, which is a typical assumption for steel and hard solids~\cite{man1996separation}. To achieve this, we can choose different values for the $\beta_j$. For example, we can choose the $\beta_j$ such that no explicit coupling of the form $\bftau \bfG$ and $\trace(\bftau \bfG)$ appears in~\eqref{eqn:coaxial-moduli}. The minimal set of equations needed  to achieve this are
\[
\beta _{15}=\beta _2, \quad \beta _{11}=\beta _4, \quad  \beta _{12}=\beta _6-\beta _{13},
\quad  \beta_{18}=\beta _7-\beta _{14}, \quad  \beta _{10} = \beta_{16} + 2\beta _{17}.
\]


\subsection{\edit{Application: measuring initial stress with shear waves}}
\label{sec:shearwaves}


\edit{For hard solids, like metals, ceramics, composites and concrete, third-order elasticity is sufficient to describe the elastic response for most scenarios encountered in industry~\cite{man_hartigs_1998,li2020ultrasonic}. The topic has also long interested the geophysics community~\cite{tolstoy1982elastic} as stress in the Earth alters the propagation of seismic waves.

Among the applications of ultrasonic waves in industry, many have tried to use ultrasonic waves to assess  the initial stress within a solid~\cite{hirao_electromagnetic_2017}. In principle, ultrasonic waves could give a quick and non-invasive way to measure the stress, as wave speeds are easily related to the stress. One significant hurdle is to separate the influence of the stress-induced anisotropy from the texture-induced anisotropy~\cite{man1996separation,hirao_electromagnetic_2017}. This challenge motivates us to develop the framework shown in this section, as it is simpler to find an answer when the elastic moduli $\boldsymbol {\mathcal A}$ are written explicitly in terms of the initial stress and texture.

We study bulk shear waves of the form
\begin{equation} \label{eqn:shearwave}
    \vec{u} = \vec{U} e^{i k( n_1 x_1 + n_2 x_2 - vt)},
\end{equation}
where $\vec{U} = (U_1,U_2,U_3)$ is the constant amplitude vector,  $\vec n = (n_1, n_2, 0)$ is the unit vector in the direction of propagation, $k$ is the wavenumber and $v$ is the speed.
The most common scenario found in hard solids is to have the texture aligned with the initial stress and so, we choose a coordinate system $(x_1,x_2,x_3)$ aligned with the principal directions of the initial stress $\bftau = \mathrm{diag}(\tau_1,\tau_2,\tau_3)$ and the direction of anisotropy $\bfM = (1,0,0)$.
We then find from \eqref{A_fourth_order} and \eqref{eqn:coaxial-moduli} that
\begin{equation} \label{eqn:moduli-coaxial}
    \mathcal A^0_{ipjq} = 0 \quad \text{unless} \quad \begin{cases}
               p=i \; \& \; q =j, \;\;\; \text{or} \\
               p=q \; \& \; i =j, \;\;\; \text{or} \\
               p=j \; \& \; q =i.
            \end{cases}
\end{equation}
See Figure~\ref{fig:nmnm} for an illustration, noting that the texture vector $\bfM$ is also aligned with the principal stress direction $\tau_1$.
Note that due to the symmetry of the Cauchy stress, we have the following connections~\cite{shams2011initial},
\begin{align}
& \mathcal A_{1212} = \tau_2 + \mathcal A_{1221}, \quad  \mathcal A_{1212} = \tau_2 + \mathcal A_{1221}, \\
& \mathcal A_{3131} = \tau_1 + \mathcal A_{1331}, \quad  \mathcal A_{3232} = \tau_2 + \mathcal A_{2332}.
\end{align}

 Substituting~\eqref{eqn:shearwave} and \eqref{eqn:moduli-coaxial} into the equation of motion \eqref{eqn:linear-motion} leads to the following eigenvalue problem~\cite{li2020ultrasonic}:
\begin{multline}
\label{eqn:governing-shear-waves}
[\mathbf Q(\vec n) - \rho v^2 \mathbf I] \vec U = \vec 0, \quad \text{where} \quad
\\
\mathbf Q(\vec n) =  \begin{bmatrix}
    \mathcal A_{1111} n_1^2 + \mathcal A_{1212}  n_2^2 &(\mathcal A_{1122} + \mathcal A_{1221})  n_1 n_2  & 0
    \\
     (\mathcal A_{1122} + \mathcal A_{1221}) n_1 n_2 &   \mathcal A_{2121} n_1^2 +  \mathcal A_{2222} n_2^2 & 0
    \\
    0 & 0 & \mathcal A_{3131} n_1^2  + \mathcal A_{3232}  n_2^2
\end{bmatrix}
\end{multline}
is the acoustic tensor.
Substituting the moduli given by combining \eqref{eqn:coaxial-moduli} and \eqref{A_fourth_order}, we then find
\begin{align} \label{eqn:moduli-aligned-texture}
    & \mathcal A_{1 1 1 1} =  \beta_1 +  2 \beta_ 3 + \beta_5 + \beta_8 + 2 \beta_9 + \tau_1 [1 + \beta_2 + \beta_4 + 2 (\beta_6 + \beta_7) + \beta_{10}]
    \notag    \\
        & \qquad \qquad+ (\tau_2 + \tau_3) (\beta_{11} + 2 \beta_{12} + \beta_{15} +
        \beta_{16} + 2 \beta_{18}),
\notag    \\
    & \mathcal A_{1 2 2 1} = [\beta_1 + \beta_3 + \tau_1 (\beta_2 + \beta_7) + \tau_2( \beta_{14} + \beta_{15} +  \beta_{18}) + \tau_3 (\beta_{15} +  \beta_{18})]/2,
\notag    \\
    & \mathcal A_{1 1 2 2} =  \beta_8 + \beta_9 + \tau_1 (\beta_4 + \beta_6) + \tau_2 (\beta_{11} +  \beta_{12} + \beta_{13} + \beta_{17}) + \tau_3 (\beta_{11} + \beta_{12}),
  \notag   \\
    & \mathcal A_{2 2 2 2} =  \beta_1 + \beta_8 + \tau_1 (\beta_2 + \beta_4)  + \tau_2(1 + \beta_{11} +  2 \beta_{13} + 2\beta_{14} + \beta_{15}) + \tau_3(\beta_{11} + \beta_{15}),
\notag    \\
    & \mathcal A_{1 3 3 1} = [\beta_1 + \beta_3 + \tau_1 (\beta_2 + \beta_7)  + \tau_2 (\beta_{15} +  \beta_{18}) + \tau_3(\beta_{14} + \beta_{15} +  \beta_{18})]/2,
\notag    \\
    & \mathcal A_{2 3 3 2} = [\beta_1 + \tau_1 \beta_2 + (\tau_2 + \tau_3) (\beta_{14} +
          \beta_{15})]/2.
\end{align}
With Equations~\eqref{eqn:moduli-aligned-texture} and \eqref{eqn:governing-shear-waves} it is now straightforward to calculate shear waves speeds and to design methods to measure the stress.

\begin{figure}[h] \label{fig:shearwaves}
    \centering
    \includegraphics[width=0.6\linewidth]{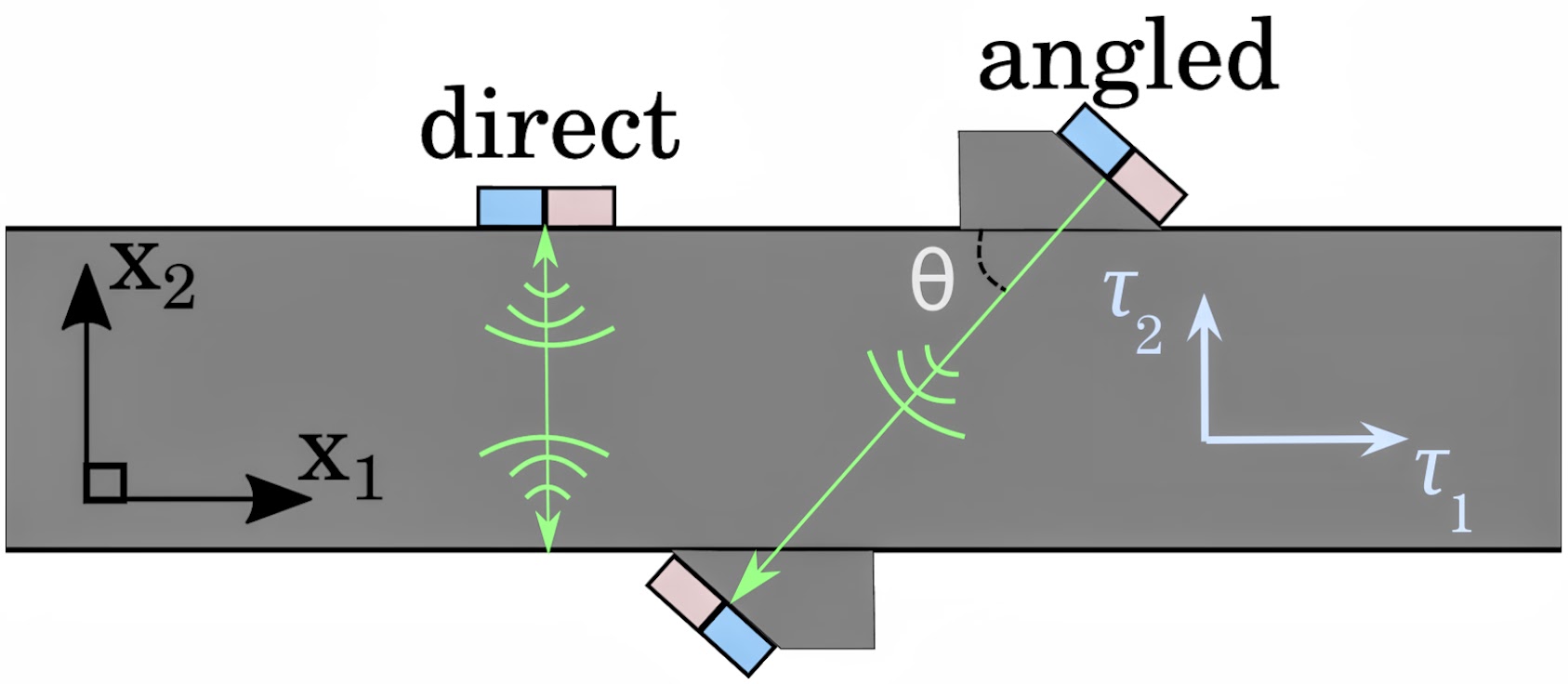}
    \caption{\edit{Two different methods to measure the stress with shear waves. The Elastic Birefringence method uses waves sent directly across the sample under stress (left). The Angled Shear Wave method uses waves sent at different angles with respect to the principal directions of stress (right).}}
\end{figure}

The most popular technique for non-destructive evaluation of stress is the \emph{Elastic Birefringence method}~\cite{tatsuo1968acoustical}.
There, two shear waves are sent directly across the stress ($n_1 = 0$, $n_2 = 1$), as shown in Figure~\ref{fig:shearwaves}.
The first wave, with speed $v_{21}$, is polarised along the first principal direction of stress ($\vec U=(U_1,0,0)$), and the second, with speed $v_{23}$, is polarised along the third principal direction of stress ($\vec U = (0,0,U_3)$).
It is then a simple exercise to show that
\begin{equation}
\rho (v_{21}^2 - v_{23}^2) = [\beta_3 + \tau_1 \beta_7 + \tau_2 \beta_{18} + \tau_3 (\beta_{18} - \beta_{14})]/2.
\end{equation}
Often, one or two directions are stress-free in many practical scenarios as, for example, is the case for railways~\cite{hirao1994advanced}.
Taking $\tau_2 = \tau_3 =0$, and assuming we know the values of $\beta_3$ and $\beta_7$, then the above formula would give us access to the stress.
However, the constant $\beta_3$ is due to texture anisotropy, as can be seen from~\eqref{eqn:coaxial-moduli}. As noted in the literature~\cite{hirao1994advanced},  texture anisotropy can vary greatly from one sample to another, and the term $\beta_3$ is often of a comparable magnitude to, say, $\tau_1 \beta_7$.
This uncertainty is one of the shortcomings of the birefringence method.

Almost all existing techniques to measure stress with ultrasonic waves suffer from the same problem as the Elastic Birefringence method: an a priori knowledge of the constants is required, but their value  can vary greatly from one sample to another, making them difficult to measure reliably~\cite{li2020ultrasonic}.
A recently proposed method does away with this problem as it is parameter-free: the \emph{Angled Shear Wave} method~\cite{li2020ultrasonic}, see Figure~\ref{fig:shearwaves} for an illustration of its working principle.
Below we show that this method can be extended to include  texture anisotropy, provided the effect of the texture is of the same order as the effect of the stress on the wave speed.

To calculate the speed of a shear wave travelling at an angle with respect to the principal directions of initial stress, we take $U_3 = 0$ in~\eqref{eqn:governing-shear-waves} and solve for $\rho v^2$, which leads to two solutions.
We identify which one corresponds to a shear wave by using \eqref{eqn:moduli-aligned-texture} and taking the limit of no stress anisotropy ($\tau_1=\tau_2=\tau_3 =0$) and no texture anisotropy ($\beta_3 = \beta_5 = \beta_9 = 0$). The shear wave speed in this limit is equal to $\sqrt{\mu/\rho}$,  noting that $\beta_1 = 2 \mu$ and $\beta_8 = \lambda$, where $\mu$ and $\lambda$ are the Lam\'e parameters.

Let $v_\theta$ denote the speed of the identified shear wave. As explained by Li et al.~\cite{li2020ultrasonic}, the formulas for these wave speeds are only accurate up to first order in the stress $\tau_j \sim \epsilon$.
In hard polycrystalline materials like steel and concrete, the influence of texture and stress on ultrasonic wave speeds are typically  both weak \cite{man1996separation}.
This implies that the texture anisotropy parameters $\beta_3, \beta_5, \beta_9$ are also small, as confirmed by several experiments \cite{egle1976measurement,hirao1994advanced}.
Assuming that $\beta_3 \sim \beta_5 \sim \beta_9 \sim \epsilon$, then we can simplify the expression for $v_\theta$ by taking a series expansion in $\epsilon$ and keeping only the first-order terms, to give
\begin{multline}
    \rho v_\theta^2 = \mu + \frac{\beta_3}{2} + \beta_5 n_1^2 n_2^2 + \frac{\tau_1}{2} (\beta_2 + \beta_7 +  2 n_1^2 + 2  \beta_{10}n_1^2n_2^2)
    \\
    + \frac{\tau_2}{2} [\beta_{14} + \beta_{15}  + \beta_{18} + 2 n_2^2  + 2 n_1^2 n_2^2 (\beta_{16} - 2 \beta_{17})]
    + \frac{\tau_3}{2} (\beta_{15}  + \beta_{18} + 2  \beta_{16} n_1^2 n_2^2).
\end{multline}
Now we take $n_1 = \cos \theta$ and $n_2 = \sin \theta$ and compute the difference $\rho (v_{\theta_1}^2 - v_{\theta_2}^2)$, where $\theta_2 = \pm\theta_1 \pm \pi/2$, to obtain
\begin{equation}
\frac{\rho (v_{\theta_1}^2 - v_{\theta_2})^2}{\cos(2 \theta_1) } = \tau_1 - \tau_2.
\end{equation}
This extension of the angled shear wave method is valid provided that both waves travel through a region with the same texture, that the texture is aligned with the stress, and that the initial stress and the texture both have a small effect on the wave speed.
It is independent of the material parameters and allows for a direct measurement of the initial stress in many real-world situations, such as that present in railways or prestressed concrete (see Figure \ref{fig:1}(c)).

}


\subsection{Linearised equation of ISRI}
\label{sec:lin}


In linear elasticity, the term $\left[\frac{\partial\bfsigma}{\partial \bfF}\right]_{\bfF=\bfI}$ is required to solve equilibrium and dynamic problems~\cite{shams2011initial}.
Here we focus on using ISRI to derive restrictions on the contraction $\left[\frac{\partial \bfsigma}{\partial \bfF}\right]_{\bfF = \bfI}: \bfepsilon$.

We follow the procedure developed in~\cite{gower2017new}. We differentiate \eqref{eqn:2nd} with respect to $\hat{\bfF}$, while holding $\tilde \bfF$, $\bftau$, and $\bfG$ fixed, to reach and contract both sides on the right with $\bfepsilon$, and obtain
\begin{equation} \label{eqn:linISRI}
	 \bftau:\bfepsilon=\trace\bfepsilon\hat{W}+\left[\frac{\partial W}{\partial \bftau}\right]_{\bfF=\bfI}:\left[\frac{\partial\bfsigma}{\partial \bfF}\right]_{\bfF=\bfI}:\bfepsilon + 2\left[\frac{\partial W}{\partial \bfG}\right]_{\bfF=\bfI} : (\bfG\bfepsilon),
\end{equation}
where we used the identity $\bftau = \left[\frac{\partial W}{\partial \bfF}\right]_{\bfF=\bfI}$.
The second and third terms in \eqref{eqn:linISRI} are expanded in terms of powers of the initial stress as follows:
\begin{align}
\label{eqn:dWdtau}
& \left[\frac{\partial W}{\partial \bftau}\right]_{\bfF=\bfI}=\beta_1\bfI+\beta_2\bftau+\beta_3\bftau^2+\beta_4\bfG+\beta_5\left<\bfG\bftau\right>_s,
\\
\label{eqn:kappas}
 & \left[\frac{\partial W}{\partial \bfG}\right]_{\bfF=\bfI}=\beta_6\bfI+\beta_4\bftau+\tfrac{1}{2}\beta_5\bftau^2.
\end{align}
Here the coefficients $\beta_1$, $\beta_2$, and $\beta_3$ are obtained by differentiating $W$ with respect to $\trace\bftau$, $\trace\bftau^2$, and $\trace\bftau^3$, respectively, and evaluating at $\bfF=\bfI$ \cite{gower2017new}.
The additional coefficients $\beta_4$, $\beta_5$, and $\beta_6$ are expressed as
\begin{align}
	\beta_4=& \left[\frac{\partial W}{\partial \bfM.\bftau\bfM}\right]_{\bfF=\bfI}= \left[ \frac{\partial W}{\partial I_{13}}\right]_{\bfF=\bfI}+\left[ \frac{\partial W}{\partial I_{14}}\right]_{\bfF=\bfI}, \notag \\[4pt]
	\beta_5=&  2\left[\frac{\partial W}{\partial \bfM.\bftau^2\bfM}\right]_{\bfF=\bfI}= 2\left[ \frac{\partial W}{\partial I_{15}}\right]_{\bfF=\bfI}+2\left[ \frac{\partial W}{\partial I_{16}}\right]_{\bfF=\bfI}, \notag\\[4pt]
	\beta_6=&\left[\frac{\partial W}{\partial \bfM.\bfC\bfM}\right]_{\bfF=\bfI}+\left[\frac{\partial W}{\partial \bfM.\bfC^2\bfM}\right]_{\bfF=\bfI}=\left[\frac{\partial W}{\partial I_4}\right]_{\bfF=\bfI}+\left[\frac{\partial W}{\partial  I_5}\right]_{\bfF=\bfI}.
\end{align}
Following \cite{gower2017new}, we use \eqref{eqn:dWdtau}, \eqref{eqn:kappas}, and \eqref{eqn:variation} to rewrite \eqref{eqn:linISRI} as
\begin{multline}
	 \trace\left(\bfepsilon\bftau\right)=\left(\hat{W}+\gamma_0\right) \trace\bfepsilon+\gamma_1 \bfM.\bfepsilon\bfM
	 +\gamma_2 \bfM.\bfepsilon\bftau\bfM
	 \\
	 + \gamma_3 \trace\left(\bfepsilon\bftau\right)
	 +\gamma_4 \trace\left(\bfepsilon\bftau^2\right) +\gamma_5\bfM.\bfepsilon\bftau^2\bfM,
	 \label{tr-relation}
\end{multline}
where  the $\gamma_i$ coefficient are expressed in terms of $\alpha_i$, $\beta_i$, and the invariants~\eqref{inv4}.
As \eqref{tr-relation} has to hold for every $\bfepsilon$, $\bftau$, and unit direction $\bfM$, we then deduce the relations
\begin{equation}  \label{eqn_with_gamma}
 \hat{W}+\gamma_0=0,\quad  \gamma_1=1,\quad\gamma_2=0,\quad\gamma_3=0,\quad\gamma_4=0,\quad\gamma_5=0.
\end{equation}
In general, solving the above equations analytically seems to be very difficult.


\subsection{Small stress $\bftau$}
\label{sec:small-stress}


Now we linearise $\left[\frac{\partial\bfsigma}{\partial \bfF}\right]_{\bfF=\bfI}$ for small $\bftau$, which  simplifies Equations~\eqref{eqn_with_gamma} and generalises results of third-order elasticity~\cite{destrade2010third, li2020ultrasonic} to include residual stresses.
%
Note that because $\bftau$ is a dimensional quantity, `small stress' means that $\|\bftau \|$ is small in comparison to $ \|\frac{\partial\bfsigma}{\partial \bfF}\|_{\bfF=\bfI, \bftau = \boldsymbol 0}$.


Previous work \cite{gower2017new} shows that to expand $\frac{\partial\bfsigma}{\partial \bfF}$ to first order in $\bftau$, we need to expand the linear ISRI~\eqref{eqn:linISRI} up to second order in $\bftau$, as follows
\begin{equation}
	\label{eqn:linear-ISRI-underbrace}
    \bftau:\bfepsilon = \trace \bfepsilon \underbrace{\hat{W}}_{1, \bftau, \bftau^2} + \Big[
		\underbrace {\frac{\partial W}{\partial \bftau}}_{1, \bftau, \bftau^2}\Big ]_{\bfF=\bfI} : \Big[ \underbrace{\frac{\partial\bfsigma}{\partial \bfF}}_{1,\bftau}\Big]_{\bfF=\bfI}:\bfepsilon  + 2\Big[ \underbrace{\frac{\partial W}{\partial \bfG}}_{1,\bftau,\bftau^2}\Big]_{\bfF=\bfI}:
         \bfG\bfepsilon.
\end{equation}
Here the terms under the braces are the different orders of $\bftau$ we need to consider. For example, ``$1, \bftau, \bftau^2$'' means we need to expand the term above up to second order in $\bftau$.

%



Returning to \eqref{tr-relation}, to use these equations and deduce restrictions for the $a_{j,i}$ we perform a series expansion up to order $\mathcal O (\bftau^2)$ of \eqref{tr-relation}, which together with equations \eqref{eqn_with_gamma}, leads to
\begin{align}
	&& \hat W + \gamma_0 = 0, && \gamma_1 = 0, && \text{up to } \mathcal O (\bftau^2),
	\label{ISRI-tau-2} \\
	&& \gamma_2 = 0, && \gamma_3 = 0, && \text{up to } \mathcal O (\bftau),
	\label{ISRI-tau-1} \\
	&& \gamma_4 = 0, && \gamma_5 = 0, && \text{for } \bftau = \boldsymbol 0.
	\label{ISRI-tau-0}
\end{align}

To   expand $\hat W$, $\partial W / \partial \bftau$, and $\partial W/\partial \bfG$, we simply expand $W$ up to third order in $\bftau$, see \cite{destrade2010third} for details, and then directly calculate $\partial W / \partial \bftau$ and $\partial W/\partial \bfG$. A third-order expansion of $W(\bfI,\bftau,\bfG)$ in $\bftau$ leads to
\begin{multline} \label{eqn:Wthird-order}
W(\bfI,\bftau,\bfG) =
   \psi _1 \trace \bftau
+ \psi _4 \trace(\bftau \bfG)
+ \psi_2 (\trace \bftau )^2
+ \psi _5 (\trace \bftau) \trace(\bftau \bfG)
\\
+ \psi _7 [\trace(\bftau \bfG)]^2
+\psi _{10} \trace (\bftau^2)
+\psi _{13} \trace(\bftau^2 \bfG)
+ \psi_3 (\trace \bftau )^3
\\
  + \psi_6 (\trace \bftau)^2 \trace(\bftau \bfG)
 +\psi_8 (\trace\bftau) \trace[(\bftau \bfG)^2]
 +\psi_9 \trace[(\bftau \bfG)^3]
 \\
 + \psi_{11} (\trace \bftau^2)  (\trace \bftau)
  + \psi _{12} (\trace \bftau^2)  \trace(\bftau \bfG)
  \\
   + \psi _{14} (\trace \bftau) \trace(\bftau^2 \bfG)
    +\psi _{15} \trace(\bftau \bfG) \trace(\bftau^2 \bfG) +\psi _{16} \trace \bftau^3,
\end{multline}
where the constants $\psi_j$ do not depend on $\bftau$ or $\bfG$. The first line has first- and second-order terms in $\bftau$, while the second and third line are all third-order terms. In the above we assumed that $W(\bfI, \bftau, \bfG) \to 0$ when $\bftau \to \boldsymbol 0$.
By differentiating the above expression with respect to  either $\bftau$ or $\bfG$, we can calculate $\frac{\partial W}{\partial \bftau}$ and $\frac{\partial W}{\partial \bfG}$. For example,
\begin{multline}
\Big[ \frac{\partial W}{\partial \bfG}\Big]_{\bfF=\bfI} \!\!\! :\bfG\bfepsilon = \psi _4 \trace(\bfepsilon \bfG \bftau) + \psi _5 (\trace\bftau) \trace(\bfepsilon \bfG \bftau) + \psi _{13} \trace(\bfepsilon \bfG\bftau^2) +2 \psi _7 \trace(\bftau \bfG) \trace(\bfepsilon \bfG\bftau),
\end{multline}
where we discarded terms that were of third order in $\bftau$ as these are not needed to expand \eqref{eqn:linear-ISRI-underbrace} up to second order in $\bftau$.

Next we substitute \eqref{eqn:alpha-series}, and \eqref{eqn:Wthird-order}, into the equations (\ref{ISRI-tau-2} -- \ref{ISRI-tau-0}). Then we can set the coefficient multiplying the terms
\[
\trace\bfepsilon, \; \trace(\bfepsilon \bfG),\; \trace(\bfepsilon \bfG \bftau), \; \trace(\bfepsilon \bftau),  \;
\trace(\bfepsilon \bftau^2), \; \trace(\bfepsilon \bfG \bftau^2)
\]
 to zero, because the equations (\ref{ISRI-tau-2} -- \ref{ISRI-tau-0}) must hold for every $\bftau$, $\bfepsilon$, and $\bfG$. \edit{We provide these equations, and their solution, as supplementary material in the form of a Mathematica~\cite{Mathematica} notebook, which is also available as a Github Gist~\cite{mathematica-gist}. They are too long to reproduce here.}
 These 24 equations can be written  in terms of the $a_{i,j}$ and $\psi_n$ coefficients only. As our goal is only to restrict the $a_{i,j}$ coefficients, we can eliminate the  $\psi_n$ coefficients and reduce the 24 equations to 7 independent equations, which can be written only in terms of the 21 $a_{i,j}$ coefficients. In this process we deduce that $\psi_1 = \psi_4 = 0$, which confirms that $W$ in Equation~\eqref{eqn:Wthird-order} has no first-order terms in $\bftau$. In conclusion, we know that there must be only 14 independent coefficients of the form $a_{i,j}$, which is exactly the same number of independent coefficients for a third-order elastic material with transverse texture anisotropy~\cite{destrade2010third}.

 When applying the restriction~\eqref{eqn:linear-ISRI-underbrace} to the incremental stress~\eqref{eqn:stress-linear-tau} we expect a response which is asymptotically equivalent to that of a third-order elastic material with transverse texture anisotropy~\cite{destrade2010third}, which has undergone a deformation corresponding to the stress field $\bftau$.
 This is because the ISRI restriction is automatically satisfied by classical hyperelastic materials~\cite{gower2017new,gower2015initial}.




Of the seven independent equations we were able to deduce for the $a_{i,j}$, the three simplest equations are
\begin{align} \label{eqn:a30}
a_{3,0} = & \frac{a_{2,0} }{a_{1,0}}\left(a_{4,0}+1\right),
\\
 \label{eqn:a40}
 a_{4,0} =& \frac{\left(a _{7,0}-a _{1,1}\right) a _{1,0}^2-\left(a _{2,0}+a _{1,2}
   \left(a _{5,0}+a _{6,0}\right)+a _{1,1} \left(3 a _{5,0}+a
   _{6,0}\right)\right) a _{1,0}-2 a _{2,0} a _{5,0}}{2 a _{2,0} a
   _{5,0}+a _{1,0} \left(a _{2,0}+2 a _{5,0}\right)},
\\
 a_{9,0} =& \frac{\left(a _{10,0}-a _{1,2}\right) a _{1,0}^2 +
a _{2,0} a _{1,0} \left(-2 a_{1,1}-2 a _{1,2}+a _{4,0}+1\right)}{a _{1,0} \left(a _{1,1}+a _{1,2}\right)}
\notag\\
 \label{eqn:a90}
 & - \frac{ \left(3 a _{1,1}+a _{1,2}+2 a_{4,0}\right) a _{1,0} a _{6,0}
+ 2 a _{2,0} \left(a _{4,0}+1\right) a _{6,0}}{a _{1,0} \left(a _{1,1}+a _{1,2}\right)}.
\end{align}
\edit{We provide all seven restrictions as supplementary material, in the form of a Mathematica notebook,  available as a Github Gist~\cite{mathematica-gist} (the other four equations are too long to reproduce here.)
The notebook also demonstrates that these restrictions do indeed solve Equations~\eqref{ISRI-tau-2} to \eqref{ISRI-tau-0}}.

\edit{Specifically we note} that Equation~\eqref{eqn:a40} reduces to the linearised ISRI condition for materials with no texture anisotropy given by Equation (4.36) in \cite{gower2017new} when taking $a_{2,0} = a_{3,0} = a_{6,0} = 0$.
To verify equations~(\ref{eqn:a30} - \ref{eqn:a90}) we checked these results directly against the instantaneous moduli of a hyperelastic material with fibre reinforcement~\cite{destrade2010third} and obtained the same equations.


\subsection{\edit{Example: texture aligned with the initial stress}}
\label{sec:ISRI-aligned-small-stress}


\edit{
Substituting the coefficients shown in \eqref{eqn:coaxial-moduli} into the ISRI condition~\eqref{eqn:linear-ISRI-underbrace} leads us to conclude that $\beta_{13}, \beta_{14}, \beta_{16}, \beta_{17}$ are dependent on the other $\beta_j$ coefficients. In particular, using (\ref{eqn:a30},\ref{eqn:a40},\ref{eqn:a90}), we can write $\beta_{13}$ and $\beta_{17}$ as
\begin{align}
   &  \beta _{13} \beta _1 = \beta _2 \beta _9+\beta _1 \left(\beta _{15}+1\right)+\beta _8 \left(\beta
   _2+2 \left(\beta _{14}+\beta _{15}+1\right)\right),
\\
   & \beta _{17} \beta _1 = 2 \left(\beta _3+\beta _9\right)+\beta _2 \left(2 \beta _3+\beta _5+\beta
  _9\right)+\left(\beta _3+2 \beta _9\right) \beta _{14}+\beta _1 \left(\beta _2-\beta
  _{15}\right)+2 \beta _9 \beta _{15}.
\end{align}
The equations for $\beta_{14}$ and $\beta_{17}$ are more complicated and are given in a supplementary Mathematica notebook, available as a Github Gist~\cite{mathematica-gist}.

Counting, we see that there are 14 independent coefficients for the moduli~\eqref{eqn:coaxial-moduli}, exactly the same number of independent coefficients needed for hyperelastic third-order elastic material~\cite{destrade2010third} when the strain tensor is co-axial with the texture direction.
}

\section{Conclusion}


\edit{There are many elastic solids where the combined influence of initial stress and texture is unavoidable. Both induce anisotropy in the material response.}
In this paper, we developed an elastic theory for strain energy functions that represent compressible materials with texture anisotropy under initial stress.
In  biological tissues these strain energies can be used to model collagen-reinforced materials under residual stress; in man-made solids, they can represent materials under stress with a preferred grain direction or reinforced/prestressed by fibres and cables.

To deduce the strain energy and the associated stress tensors, we used a restriction called initial stress reference independence (ISRI).
In simple terms, this restriction implies that the strain energy density, per unit mass of the reference, depends only on the deformation gradient, initial stress tensor, and texture directions. \edit{Strain energy functions that do not satisfy ISRI can lead to non-physical material responses \cite{gower2017new}.}
Here we investigated both the nonlinear and linearized forms of the strain and strain energy.

The present study provides a detailed extension of ISRI as developed by Gower and collaborators~\cite{gower2017new,gower2015initial} to a more general case including transverse texture anisotropy.
We developed two examples of strain energy densities for compressible initially stressed transversely isotropic materials that satisfy ISRI.
We also investigated the inflation of residually-stressed compressible transversely isotropic tubes, \edit{using a residual stress field which captures what is observed in many biological tubular structures. 
We then studied the tension of a welded steel plate using the developed model, and found a physically acceptable behaviour.
For both problems, we observed a considerable influence of the initial stress on the Cauchy stress distribution.
}

\edit{Finally, the linearised theory developed in this paper
can be used to investigate small-amplitude wave propagation or vibration analysis in initially-stressed solids with transverse texture.
We used this linearisation to propose a non-destructive method to measure initial stress directly with shear waves, a process which is needed, and suited to, hard solids such as metals~\cite{kube2015acoustoelasticity, man1996separation,li2020ultrasonic}. }


\section*{Acknowledgments}

\edit{SM thanks Dr Raja Chakaraborti, Tripura Institute of Technology, Agartala, India, for an enlightening discussion on Masubuchi and Martin's solution \cite{masubuchi1966investigation} for a welded plate.}
The research of MD was supported by a 111 Project for International Collaboration (Chinese Government, PR China) No. B21034 and by a grant from the Seagull Program (Zhejiang Province, PR China).
AG gratefully acknowledges partial funding provided by IN2TRACK3 - European Commission - Horison Europe and Network Rail LTD and FRA - Federal Rail Association of America (FR21RPD31000000039).


\appendix


\section{Determining the second strain energy as a function of initial stress and preferred direction}
\label{AppendixA}

\edit{
Here we express the strain energy potential \eqref{w-2}:
\begin{equation}
 W= \frac{\mu_1}{2J_0}\left(\trace(\bfB_0\bfC) - 3 \right) + \frac{\mu_2}{2J_0} \left(\bfM.\bfC\bfB_0\bfC\bfM-1\right)-\frac{\mu_1}J_0\frac{(1-\left(JJ_0\right)^{-\beta})}{\beta},\label{sefc}
 \end{equation}
as a function of initial stress and texture. Presently, this strain energy is expressed as a function of the initial strain $\bfB_0$.
But the evaluation of $\bfB_0$ requires the knowledge of the stress-free configuration $\mathcal{R}_0$ which is usually unattainable. Hence, we aim to replace $\bfB_0$ with a known function of the initial stress $\bftau$ and of the preferred direction $\bfM$.
This step necessitates the inversion of a non-linear anisotropic constitutive relation.}

\edit{The initial stress is obtained by substituting $\bfF=\bfI$ in the constitutive relation \eqref{s2}, as
\begin{equation}
    \bftau= \frac{\mu_1}J_0\bfB_0+\frac{\mu_2}J_0 \left(\bfB_0\bfM\otimes\bfM+\bfM\otimes\bfM\bfB_0\right) - \frac{\mu_1} {J_0^{\beta+1}}\bfI.\label{tau_1}
\end{equation}
Note that $\bfM$ and $\bfM_0$ are the preferred directions in the stressed and the stress-free reference configurations, respectively, where $\bfM=\bfF_0\bfM_0$.}

\edit{To find $\bfB_0$ in terms of $\bftau$ and $\bfM$, we need to invert the relation \eqref{tau_1}.
To this end, \eqref{tau_1} is expressed in the form
\begin{equation}
  J_0\left(\bftau +  \frac{\mu_1} {{J_0}^{\beta+1}} \bfI\right)=\left(\mu_1\mathbb{I} +\mu_2\bfI\boxtimes\bfM\otimes\bfM+\mu_2\bfM\otimes\bfM\boxtimes \bfI\right) \bfB_0, \label{invert1}
\end{equation}
where we define the fourth-order identity tensor as $\mathbb{I}=\bfI\boxtimes\bfI$ and the square tensor product by $\left(\bfA\boxtimes\bfB\right)\bfC = \bfA\bfC\bfB^{T}$ for any second-order tensors  $\bfA$, $\bfB$, $\bfC$.
We invert \cite{jog2006derivatives} the fourth-order tensor $\left[\mu_1\mathbb{I} +\mu_2\bfI\boxtimes\bfM\otimes\bfM+\mu_2\bfM\otimes\bfM\boxtimes \bfI\right]$ and multiply the inverse on both the sides of \eqref{invert1} to express $\bfB_0$  as a function of $\bftau$ and $\bfM$:
\begin{equation}
  \bfB_0=a_1\hat{\bftau}+a_2 \hat{\bftau}\bfM\otimes\bfM+ a_2\bfM\otimes\bfM\hat{\bftau}+a_3\bfM\otimes\bfM\hat{\bftau}\bfM\otimes\bfM,
  \label{invert2}
\end{equation}
where  $\hat{\bftau}=J_0\left(\bftau +  \frac{\mu_1} {{J_0}^{\beta+1}}\bfI\right)$ and the scalar parameters $a_1$, $a_2$, $a_3$ are calculated as
\begin{equation}
    a_1=\frac{1}{\mu_1},
    \qquad
    a_2 = - \frac{\mu_2}{\mu_1(\mu_1 + \mu_2 I_{\bfM})},
    \qquad
    a_3 = \frac{2 \mu _2^2}{\mu_1 \left(\mu_1 ^2+3 \mu_1  \mu_2 I_{\bfM} + 2 \mu _2^2 I_{\bfM}^2 \right)},
\end{equation}
by substituting \eqref{invert2} in \eqref{tau_1} and thereafter equating the coefficients of $\hat{\bftau}$, $\left(\hat{\bftau}\bfM\otimes\bfM+\bfM\otimes\bfM \hat{\bftau}\right)$, and $\bfM\otimes\bfM\hat{\bftau} \bfM\otimes\bfM$ on both sides of the resulting equation.}

\edit
{Finally, we use the derived expression of $\bfB_0$ \eqref{invert2} to express the strain energy density \eqref{sefc} as a function of initial stress, as
\begin{align}
    W= &\frac{1}{2}\left(I_9 + \frac{\mu_1} {{J_0}^{\beta+1}}I_1 - 3\mu_1\right)
    + \frac{\mu_1}{2} \left[2 a_2 \left(I_{13} + \frac{\mu_1} {{J_0}^{\beta+1}}I_4\right) + a_3 \left(I_{\bftau_4} I_4 + \frac{\mu_1} {{J_0}^{\beta+1}}I_4I_{\bfM}\right)\right]
    \nonumber
    \\
    & + \frac{\mu_2}{2} \left[a_1 \bfM \cdot \bfC\bftau\bfC\bfM + 2a_2I_4I_{13} + a_3I_4^2I_{\bftau_4} + \frac{\mu_1} {{J_0}^{\beta+1}}\left(a_1 I_5 + 2 a_2 I_4^2 + a_3 I_4^2 I_{\bfM}\right) - 1\right]
    \nonumber\\
    &-{\mu_1}\frac{(1-\left(JJ_0\right)^{-\beta})}{\beta J_0},
\end{align}
which includes most invariants developed in \eqref{inv4}, e.g., $I_1$, $I_3$, $I_{\bfM}$ $I_4$, $I_5$, $I_{\bftau_4}$, $I_9$, $I_{13}$, and $\bfM \cdot\bfC\bftau\bfC\bfM$.
However, it also involves $J_0$, which is an invariant of the (unknown) initial strain.
To express $J_0$ in terms of the known initial stress invariants, we evaluate the determinant on both sides of the equation
\begin{equation}
 J_0\left(\bftau-g(J_0)\bfI\right)=  {\mu_1}\bfB_0+{\mu_2} \left(\bfB_0\bfM\otimes\bfM+\bfM\otimes\bfM\bfB_0\right).\label{find_det}
\end{equation}
The determinant of the right-hand side of \eqref{find_det} can be calculated as
 \begin{align}
     \text{det}\left[{\mu_1}\bfB_0+{\mu_2} \left(\bfB_0\bfM\otimes\bfM+\bfM\otimes\bfM\bfB_0\right)\right]
     &=J_0^2\text{det}\left[{\mu_1}\bfI+{\mu_2} \left(\bfC_0\bfM_0\otimes\bfM_0+\bfM_0\otimes\bfM_0\bfC_0\right)\right]\nonumber\\
     &=J_0^2\left(\mu_1^3+\mu_1^2I_{\bfP_1}+\mu_1 I_{\bfP_2}+ I_{\bfP_3}\right),
     \label{dt1}
 \end{align}
where $I_{\bfP_1}=\bfM \cdot \bfM$, $I_{\bfP_2}=\half(\bfM \cdot \bfM)^2 + \bfM_0 \cdot \bfC_0^2\bfM_0$, and $I_{\bfP_3}=0$.
Following similar steps, we also expand the determinant of left-hand side of \eqref{find_det} to finally obtain
 \begin{multline}
     g(J_0)^3-g(J_0)^2I_{\bftau_1}+g(J_0) I_{\bftau_2}- I_{\bftau_3}
     \\
     +\frac{1}{J_0}\left[\mu_1^3+\mu_1^2\mu_2 (\bfM \cdot \bfM) + \mu_1\mu_2^2 (\bfM_0 \cdot \bfC_0^2\bfM_0) + \half\mu_1\mu_2^2(\bfM \cdot \bfM)^2\right] = 0.
     \label{solveit}
 \end{multline}
This equation for $J_0$ still contains another invariant, $\bfM_0 \cdot \bfC_0^2\bfM_0$, of the initial strain, because  $\vec M_0$ and $\bfC_0$ are both associated with the {unknown} stress-free configuration.
Hence, an additional equation correlating the invariants of initial stress and strain is required. This is accomplished by multiplying $\bfM\otimes\bfM$ on both sides of \eqref{find_det} and calculating the trace. With some simplifications, we obtain this additional equation as
 \begin{equation}
     J_0\left[\bfM \cdot \bftau\bfM-g\left(J_0\right)(\bfM \cdot \bfM)\right]=\mu_1\bfM_0.\bfC_0^2\bfM_0 +2\mu_2 \left(\bfM_0.\bfC_0^2\bfM_0\right)(\bfM \cdot \bfM),
     \end{equation}
     which implies that
     \begin{equation}
     \bfM_0\cdot \bfC_0^2\bfM_0 = \frac{J_0\left[I_{\bftau_4} - g\left(J_0\right)(\bfM \cdot \bfM)\right]}{\mu_1+2\mu_2(\bfM \cdot \bfM) }.\label{mc2m}
 \end{equation}
 We can substitute \eqref{mc2m} into \eqref{solveit} and  solve to express $J_0$ completely in terms of the invariants of initial stress $\bftau$.
 Hence, we finally obtain a strain energy function where all terms and parameters are completely expressed with the quantities associated with the stressed reference. We can similarly describe Cauchy stress completely from the stressed reference by substituting $\bfB_0$ \eqref{invert2} and $J_0$ in \eqref{s2}.}

\bibliographystyle{plain}
\bibliography{hyper1}

\begin{thebibliography}{10}

\bibitem{agosti2018constitutive}
A.~Agosti, A.~L. Gower, and P.~Ciarletta.
\newblock The constitutive relations of initially stressed incompressible
  {M}ooney-{R}ivlin materials.
\newblock {\em Mechanics Research Communications}, 93:4--10, 2018.

\bibitem{anand2020continuum}
L.~Anand and S.~Govindjee.
\newblock {\em Continuum Mechanics of Solids}.
\newblock Oxford University Press, 2020.

\bibitem{choudhury2020temperature}
S.~Choudhury, T.~Medhi, D.~Sethi, S.~Kumar, B.~Saha Roy, and S.~C. Saha.
\newblock Temperature distribution and residual stress in friction stir welding
  process.
\newblock {\em Materials Today: Proceedings}, 26:2296--2301, 2020.

\bibitem{ciarletta2016residual}
P.~Ciarletta, M.~Destrade, and A.~L. Gower.
\newblock On residual stresses and homeostasis: an elastic theory of functional
  adaptation in living matter.
\newblock {\em Scientific reports}, 6:24390, 2016.

\bibitem{ciarletta2016morphology}
P.~Ciarletta, M.~Destrade, A.~L. Gower, and M.~Taffetani.
\newblock Morphology of residually stressed tubular tissues: Beyond the elastic
  multiplicative decomposition.
\newblock {\em Journal of the Mechanics and Physics of Solids}, 90:242--253,
  2016.

\bibitem{destrade2001explicit}
M.~Destrade.
\newblock The explicit secular equation for surface acoustic waves in
  monoclinic elastic crystals.
\newblock {\em The Journal of the Acoustical Society of America},
  109(4):1398--1402, 2001.

\bibitem{destrade2010third}
M.~Destrade, M.~D. Gilchrist, and R.~W. Ogden.
\newblock Third-and fourth-order elasticities of biological soft tissues.
\newblock {\em The Journal of the Acoustical Society of America},
  127(4):2103--2106, 2010.

\bibitem{destrade2010third1}
M.~Destrade and R.~W. Ogden.
\newblock On the third-and fourth-order constants of incompressible isotropic
  elasticity.
\newblock {\em The Journal of the Acoustical Society of America},
  128(6):3334--3343, 2010.

\bibitem{destrade2013stress}
M.~Destrade and R.~W. Ogden.
\newblock On stress-dependent elastic moduli and wave speeds.
\newblock {\em The IMA Journal of Applied Mathematics}, 78(5):965--997, 2013.

\bibitem{du2018modified}
Y.~Du, C.~L{\"u}, W.~Chen, and M.~Destrade.
\newblock Modified multiplicative decomposition model for tissue growth: Beyond
  the initial stress-free state.
\newblock {\em Journal of the Mechanics and Physics of Solids}, 118:133--151,
  2018.

\bibitem{du2019influence}
Y.~Du, C.~L{\"u}, M.~Destrade, and W.~Chen.
\newblock Influence of initial residual stress on growth and pattern creation
  for a layered aorta.
\newblock {\em Scientific Reports}, 9(1):1--9, 2019.

\bibitem{du2019prescribing}
Y.~Du, C.~L{\"u}, C.~Liu, Z.~Han, J.~Li, W.~Chen, S.~Qu, and M.~Destrade.
\newblock Prescribing patterns in growing tubular soft matter by initial
  residual stress.
\newblock {\em Soft matter}, 15(42):8468--8474, 2019.

\bibitem{egle1976measurement}
DM~Egle and DE~Bray.
\newblock Measurement of acoustoelastic and third-order elastic constants for
  rail steel.
\newblock {\em The journal of the Acoustical Society of America},
  60(3):741--744, 1976.

\bibitem{gower2015initial}
A.~L. Gower, P.~Ciarletta, and M.~Destrade.
\newblock Initial stress symmetry and its applications in elasticity.
\newblock {\em Proceedings of the Royal Society A: Mathematical, Physical and
  Engineering Sciences}, 471(2183):20150448, 2015.

\bibitem{gower2017new}
A.~L. Gower, T.~Shearer, and P.~Ciarletta.
\newblock A new restriction for initially stressed elastic solids.
\newblock {\em The Quarterly Journal of Mechanics and Applied Mathematics},
  70(4):455--478, 2017.

\bibitem{mathematica-gist}
AL~Gower.
\newblock A mathematica notebook for section 4.
\newblock {\em
  \url{https://gist.github.com/arturgower/087a1412cef6ba45c067ef51a6550dfc}},
  2022.

\bibitem{grine2021initially}
F.~Grine.
\newblock {\em Initially-stressed hyperelastic materials: Modeling, mechanical
  and numerical analysis of singular problems and identification of residual
  stress}.
\newblock PhD thesis, Universit{\'e} de Lyon; {\'E}cole nationale
  d'ing{\'e}nieurs de Tunis (Tunisie), 2021.

\bibitem{haughton1997eversion}
D.~M. Haughton and A.~Orr.
\newblock On the eversion of compressible elastic cylinders.
\newblock {\em International journal of solids and structures},
  34(15):1893--1914, 1997.

\bibitem{hirao_electromagnetic_2017}
M.~Hirao and H.~Ogi.
\newblock {\em Electromagnetic Acoustic Transducers: Noncontacting Ultrasonic
  Measurements using {EMATs}}.
\newblock Springer Series in Measurement Science and Technology. Springer
  Japan, 2017.

\bibitem{hirao1994advanced}
M~Hirao, H~Ogi, and H~Fukuoka.
\newblock Advanced ultrasonic method for measuring rail axial stresses with
  electromagnetic acoustic transducer.
\newblock {\em Research in Nondestructive Evaluation}, 5(3):211--223, 1994.

\bibitem{hoger1985residual}
A.~Hoger.
\newblock On the residual stress possible in an elastic body with material
  symmetry.
\newblock {\em Archive for Rational Mechanics and Analysis}, 88(3):271--289,
  1985.

\bibitem{hoger1993constitutive}
A.~Hoger.
\newblock The constitutive equation for finite deformations of a transversely
  isotropic hyperelastic material with residual stress.
\newblock {\em Journal of elasticity}, 33(2):107--118, 1993.

\bibitem{hoger1994positive}
A.~Hoger.
\newblock Positive definiteness of the elasticity tensor of a residually
  stressed material.
\newblock {\em Journal of elasticity}, 36(3):201--226, 1994.

\bibitem{hoger1996elasticity}
A.~Hoger.
\newblock The elasticity tensor of a transversely isotropic hyperelastic
  material with residual stress.
\newblock {\em Journal of elasticity}, 42(2):115--132, 1996.

\bibitem{Mathematica}
Wolfram~Research{,} Inc.
\newblock Mathematica, {V}ersion 12.3.1.
\newblock Champaign, IL, 2021.

\bibitem{jog2006derivatives}
C.~S. Jog.
\newblock Derivatives of the stretch, rotation and exponential tensors in
  n-dimensional vector spaces.
\newblock {\em Journal of Elasticity}, 82(2):175--192, 2006.

\bibitem{johnson1993dependence}
B.~E. Johnson and A.~Hoger.
\newblock The dependence of the elasticity tensor on residual stress.
\newblock {\em Journal of Elasticity}, 33(2):145--165, 1993.

\bibitem{johnson1995use}
B.~E. Johnson and A.~Hoger.
\newblock The use of a virtual configuration in formulating constitutive
  equations for residually stressed elastic materials.
\newblock {\em Journal of Elasticity}, 41(3):177--215, 1995.

\bibitem{kube2015acoustoelasticity}
C.~M. Kube, A.~Arguelles, and J.~A. Turner.
\newblock On the acoustoelasticity of polycrystalline materials.
\newblock {\em The Journal of the Acoustical Society of America},
  138(3):1498--1507, 2015.

\bibitem{li2020ultrasonic}
G.-Y. Li, A.~L. Gower, and M.~Destrade.
\newblock An ultrasonic method to measure stress without calibration: The
  angled shear wave method.
\newblock {\em The Journal of the Acoustical Society of America},
  148(6):3963--3970, 2020.

\bibitem{liu2020growth}
Congshan Liu, Yangkun Du, Chaofeng L{\"u}, and Weiqiu Chen.
\newblock Growth and patterns of residually stressed core--shell soft sphere.
\newblock {\em International Journal of Non-Linear Mechanics}, 127:103594,
  2020.

\bibitem{man_hartigs_1998}
C.-S. Man.
\newblock Hartig's law and linear elasticity with initial stress.
\newblock {\em Inverse Problems}, 14(2):313, 1998.

\bibitem{man1996separation}
C.~S. Man and R.~Paroni.
\newblock On the separation of stress-induced and texture-induced birefringence
  in acoustoelasticity.
\newblock {\em Journal of elasticity}, 45(2):91--116, 1996.

\bibitem{masubuchi1966investigation}
K.~Masubuchi and D.~C. Martin.
\newblock Investigation of residual stresses in steel weldments.
\newblock Technical report, Nat'l Academy of Sciences, Nat'l Research Council,
  1966.

\bibitem{merodio2016extension}
J.~Merodio and R.~W. Ogden.
\newblock Extension, inflation and torsion of a residually stressed circular
  cylindrical tube.
\newblock {\em Continuum Mechanics and Thermodynamics}, 28(1-2):157--174, 2016.

\bibitem{merodio2013influence}
J.~Merodio, R.~W. Ogden, and J.~Rodr{\'\i}guez.
\newblock The influence of residual stress on finite deformation elastic
  response.
\newblock {\em International Journal of Non-Linear Mechanics}, 56:43--49, 2013.

\bibitem{mukherjee2022constitutive}
S.~Mukherjee.
\newblock Constitutive relation, limited stretchability, and stability of
  residually stressed gent materials.
\newblock {\em Mechanics Research Communications}, 120:103850, 2022.

\bibitem{mukherjee2022influence}
S.~Mukherjee.
\newblock Influence of residual stress in failure of soft materials.
\newblock {\em Mechanics Research Communications}, 123:103903, 2022.

\bibitem{mukherjee2021generalized}
S.~Mukherjee and A.~K. Mandal.
\newblock A generalized strain energy function using fractional powers:
  Application to isotropy, transverse isotropy, orthotropy, and residual stress
  symmetry.
\newblock {\em International Journal of Non-Linear Mechanics}, 128:103617,
  2021.

\bibitem{mukherjee2021static}
S.~Mukherjee and A.~K. Mandal.
\newblock Static and dynamic characteristics of a compound sphere using initial
  stress reference independence.
\newblock {\em International Journal of Non-Linear Mechanics}, 136:103787,
  2021.

\bibitem{nye1985physical}
J.~F. Nye et~al.
\newblock {\em Physical properties of crystals: their representation by tensors
  and matrices}.
\newblock Oxford university press, 1985.

\bibitem{ogden2011propagation}
R.~Ogden and B.~Singh.
\newblock Propagation of waves in an incompressible transversely isotropic
  elastic solid with initial stress: Biot revisited.
\newblock {\em Journal of Mechanics of Materials and Structures},
  6(1):453--477, 2011.

\bibitem{raghavan2004three}
M.~L. Raghavan, S.~Trivedi, A.~Nagaraj, D.~D. McPherson, and K.~B. Chandran.
\newblock Three-dimensional finite element analysis of residual stress in
  arteries.
\newblock {\em Annals of biomedical engineering}, 32(2):257--263, 2004.

\bibitem{rivlin1949large}
R.~S. Rivlin.
\newblock Large elastic deformations of isotropic materials. v. the problem of
  flexure.
\newblock {\em Proceedings of the Royal Society of London. Series A.
  Mathematical and Physical Sciences}, 195(1043):463--473, 1949.

\bibitem{shams2011initial}
M.~Shams, M.~Destrade, and R.~W. Ogden.
\newblock Initial stresses in elastic solids: constitutive laws and
  acoustoelasticity.
\newblock {\em Wave Motion}, 48(7):552--567, 2011.

\bibitem{shariff2021anisotropic}
M.~H. B.~M. Shariff.
\newblock Anisotropic stress softening of residually stressed solids.
\newblock {\em Proceedings of the Royal Society A}, 477(2252):20210289, 2021.

\bibitem{shariff2017spectral}
M.~H. B.~M. Shariff, R.~Bustamante, and J.~Merodio.
\newblock On the spectral analysis of residual stress in finite elasticity.
\newblock {\em IMA Journal of Applied Mathematics}, 82(3):656--680, 2017.

\bibitem{spencer1971part}
A.~J.~M. Spencer.
\newblock {T}heory of invariants.
\newblock In A.C. Eringen, editor, {\em Continuum Physics}, volume~1, chapter
  Part {III}, pages 239--353. Academic Press, 1971.

\bibitem{taber2001stress}
L.~A. Taber and J.~D. Humphrey.
\newblock Stress-modulated growth, residual stress, and vascular heterogeneity.
\newblock {\em J. Biomech. Eng.}, 123(6):528--535, 2001.

\bibitem{tatsuo1968acoustical}
Tokuoka Tatsuo and Iwashimizu Yukio.
\newblock Acoustical birefringence of ultrasonic waves in deformed isotropic
  elastic materials.
\newblock {\em International Journal of Solids and Structures}, 4(3):383--389,
  1968.

\bibitem{thompson1986angular}
R.~B. Thompson, S.~S. Lee, and J.~F. Smith.
\newblock Angular dependence of ultrasonic wave propagation in a stressed,
  orthorhombic continuum: Theory and application to the measurement of stress
  and texture.
\newblock {\em The Journal of the Acoustical Society of America},
  80(3):921--931, 1986.

\bibitem{tolstoy1982elastic}
I~Tolstoy.
\newblock On elastic waves in prestressed solids.
\newblock {\em Journal of Geophysical Research: Solid Earth},
  87(B8):6823--6827, 1982.

\bibitem{vandiver2009differential}
R.~Vandiver and A.~Goriely.
\newblock Differential growth and residual stress in cylindrical elastic
  structures.
\newblock {\em Philosophical Transactions of the Royal Society A: Mathematical,
  Physical and Engineering Sciences}, 367(1902):3607--3630, 2009.

\bibitem{zheng1994theory}
Q.~S. Zheng.
\newblock Theory of representations for tensor functions -- {A} unified
  invariant approach to constitutive equations.
\newblock {\em Applied Mechanics Reviews}, 47:545--587, 1994.

\end{thebibliography}

\end{document}